\documentclass[11pt]{article}

\usepackage{epsfig,amsmath,latexsym,amssymb}
\usepackage{graphicx}
\usepackage{lscape}
\usepackage{picture, eso-pic, tikz} 
\usepackage{dsfont}
\usepackage{listings}
\usepackage{changes}
\usepackage[
breaklinks,
colorlinks=true,
pagebackref=false,
pdfauthor={},%
pdftitle={},%
citecolor=blue,
linkcolor=blue,
urlcolor=blue,
filecolor=blue
]{hyperref}      
\usepackage{cleveref}

\renewcommand{\baselinestretch}{1.1}
\def\R{{\mathbb R}}  
\def\N{{\mathbb N}}  
\def\Y{{\mathbb Y}}  
\def\A{{\mathbb A}}  
\def\E{{\mathbb E}}  %
\def\F{{\mathcal F}}  

\DeclareMathOperator*{\argmin}{arg\,min}
\newcommand{\ES}{\mathrm{ES}}
\newcommand{\one}{\mathds{1}}
\newcommand{\interior}{\operatorname{int}}
\newcommand{\conv}{\operatorname{conv}}

\newcommand{\Remm}[1]{}
\newtheorem{theo}{Theorem}[section]
\newtheorem{lemma}[theo]{Lemma}

\newtheorem{cor}[theo]{Corollary}
\newtheorem{defi}[theo]{Definition}
\newtheorem{model ass}[theo]{Model Assumptions}
\newtheorem{ass}[theo]{Assumption}

\newtheorem{rem}[theo]{Remark}
\def\EndProof{\hfill {\scriptsize $\Box$}}

\numberwithin{equation}{section}

\definecolor{MyGray}{rgb}{0.92,0.92,0.92}


\definecolor{British racing}{rgb}{0.0, 0.5, 0.0}
\def\bx{\boldsymbol{x}}

\def\bz{\boldsymbol{z}}
\def\bw{\boldsymbol{w}}

\def\bX{\boldsymbol{X}}
\def\b0{\boldsymbol{0}}

\def\bvartheta{\boldsymbol{\vartheta}}

\def\bbeta{\boldsymbol{\beta}}

\def\b0{\boldsymbol{0}}

\usepackage{todonotes}
\newcommand{\Comments}{1}
\newcommand{\mynote}[2]{\ifnum\Comments=1\textcolor{#1}{#2}\fi}
\newcommand{\mytodo}[2]{\ifnum\Comments=1%
  \todo[linecolor=#1!80!black,backgroundcolor=#1,bordercolor=#1!80!black]{#2}\fi}

\ifnum\Comments=1               
\fi

\ifnum\Comments=1               
\fi

\ifnum\Comments=1               
\fi

\begin{document}
\author{Tobias Fissler\footnote{Corresponding Author. Institute for Statistics and Mathematics, Department of Finance, Accounting and Statistics, Vienna University of Economics and Business (WU), Welthandelsplatz 1, 1020 Vienna, Austria, email: \href{mailto:tobias.fissler@wu.ac.at}{tobias.fissler@wu.ac.at}}
\and
Michael Merz\footnote{Faculty of Business Administration, University of Hamburg, Von-Melle-Park 5, 20146 Hamburg, Germany, email: \href{mailto:michael.merz@uni-hamburg.de}{michael.merz@uni-hamburg.de}} 
\and Mario V.~W\"uthrich\footnote{RiskLab, Department of Mathematics, ETH Zurich, 8092 Zurich, Switzerland, email:
\href{mailto:mario.wuethrich@math.ethz.ch}{mario.wuethrich@math.ethz.ch}}}

\date{Version of \today}
\title{Deep Quantile and Deep Composite Model Regression\footnote{We are grateful to Timo Dimitriadis for valuable discussions.}}
\maketitle

\begin{abstract}
\noindent  
A main difficulty in actuarial claim size modeling is that there is no simple off-the-shelf distribution that simultaneously
provides a good distributional model for the main body and the tail of the data. 
In particular, covariates may have different effects for small and for large claim sizes.
To cope with this problem, we introduce a deep composite regression model whose splicing point is given in terms of a quantile of the conditional claim size distribution rather than a constant.
To facilitate M-estimation for such models, we introduce and characterize the class of strictly consistent scoring functions for the triplet consisting a quantile, as well as the lower and upper expected shortfall beyond that quantile.
In a second step, this elicitability result is applied to fit deep neural network regression models.
We demonstrate the applicability of our approach and its superiority over classical approaches on a real accident insurance data set.


~


~

\end{abstract}

\section{Introduction}
\subsection{Background}

In actuarial modeling, we typically use regression models to estimate expected values 
of claim sizes exploiting systematic effects of features or covariates.
These estimated expected values are used for insurance
pricing, claims forecasting and claims reserving. The most commonly used regression model
 is the generalized linear model (GLM) with the exponential dispersion
family (EDF) as the underlying distributional model. Parameter estimation
within this EDF-GLM framework is performed by maximum likelihood estimation (MLE).
In this setup, MLE is equivalent to minimizing the corresponding deviance loss function,
 the latter being
a strictly consistent scoring function for mean.
Strictly consistent scoring functions facilitate M-estimation and forecast evaluation; further details are provided below.
For actuarial problems, there are three main issues with this approach:
\begin{itemize}
\item[(1)] The EDF only contains light-tailed distribution functions
such as the Gaussian, the Poisson, the gamma or the inverse Gaussian distributions. For this reason, the classical
EDF-GLM framework is not suitable if we have a mixture of light-tailed and heavy-tailed data. This fact is often disregarded in practice and the EDF-GLM framework is used
nevertheless. This may result in over-fitting to large claims and non-robustness in model estimation.
\item[(2)] The GLM regression structure may not be suitable or covariate engineering may be too difficult
to bring the regression problem into a GLM form. As a result, not all systematic effects such as interactions
may be correctly
incorporated into the GLM. 
\item[(3)] The GLM is usually used to estimate expected values. From a risk management perspective, one is also interested
in other quantities such as quantiles and expected shortfalls (ES). If the aforementioned distributional model
within the EDF is not suitable for the description of the observed
data, then a very accurately estimated expected value is not directly helpful in estimating quantiles and
ES.
\end{itemize}
Our paper tackles the above points;
we first discuss the related literature and
then we are going to describe our proposal.
To deal with the problem of the lack of any simple off-the-shelf distribution that fits
the entire range of the data, i.e., the body and tail of the data, one often either uses a composite (splicing) model
 or a mixture model. These two approaches can be fitted with the expectation-maximization
(EM) algorithm. Composite models are studied in 
Cooray--Ananda \cite{CoorayAnanda}, Scollnik \cite{Scollnik}, Pigeon--Denuit 
\cite{PigeonDenuit}, Gr\"un--Miljkovic \cite{Grun} and Parodi \cite{Parodi}.
These papers do not consider
a regression structure and covariates. The only actuarial publications
that we are aware of considering composite regression models are Gan--Valdez \cite{GanValdez} and Laudag\'e et al.~\cite{Laudage}. Both proposals choose very specific distributional assumptions above and below a given
threshold (splicing point) to obtain analytical tractability.
 In our proposal, we do not fix the threshold itself, but a quantile
level instead. 
This allows us for direct model fitting below and above that quantile, and, moreover, we can have a flexible regression
structure for both the body and the tail of the data. 
For mixture models, not considered in this paper, we 
refer to Fung et al.~\cite{Fung2021} and the literature therein. Both approaches, composite and mixture
models, typically use the EM algorithm for model fitting. Our approach is based on the (simpler)
stochastic gradient descent (SGD) algorithm.

Focusing on point (2) from above, there are several ways in extending (or modifying) a GLM.
These include, e.g., generalized additive models (GAMs), regression trees and tree boosting, or feed-forward
neural (FN) network regression models; we refer to Hastie et al.~\cite{HTF}. Here, we focus on FN network regression models as a straightforward
and powerful extension of GLMs. Moreover, we see that an FN network architecture can easily be designed
so that it can simultaneously fulfill different estimation tasks. This is a crucial property 
that we need in our derivations, as we jointly
estimate different regression models for the body and tail of the data.

Concerning point (3) from above, quantile regression has gained quite some
popularity in the machine learning community  to quantify uncertainty,
see Meinshausen \cite{Meinshausen} and Takeuchi et al.~\cite{Takeuchi}. 
Quantile regression has been introduced by Koenker--Bassett \cite{KoenkerBassett}, and it is widely used in statistical modeling these days; for a recent monograph we refer to
Uribe--Guillen \cite{UribeGuillen}. Our proposal extends the approach of Richman \cite{Richman3}
who combines joint estimation of a quantile and the mean within an FN network architecture, 
see Listing 8 in Richman \cite{Richman3}.

\subsection{Proposed method}
We consider two different, though closely related problems in this paper. On the one hand, we extend the work of Richman \cite{Richman3} by proposing an FN network architecture
that allows for consistent multiple quantile regression respecting monotonicity of the quantiles at different levels. 
On the other hand, this FN network architecture is at the basis of the extension of 
the joint quantile and ES regression considered in 
Guillen et al.~\cite{Guillen2021}. 
The main building blocks in our estimation procedure are strictly consistent scoring functions for our target functionals. 
These are functions of the estimated model prediction and the response variable which are minimised in expectation by the correctly specified model.
Strictly consistent scoring functions are at the core of consistent M-estimation (where M stands for minimisation) in a regression setup, 
see Dimitriadis et al.~\cite{DimitriadisFisslerZiegel2020}.
If a target functional admits a strictly consistent scoring function, it is called elicitable.
The elicitability of the mean and the quantile thus allow for mean and quantile regression, using the squared loss (or a Bregman divergence) or the pinball loss (or a generalized piecewise linear loss), respectively.
In model selection and, more generally, forecast ranking,  Gneiting \cite{Gneiting} and Gneiting--Raftery \cite{GneitingRaftery} advocate for the usage of strictly consistent scoring functions since they honour honest and truthful forecasts.


In a first step, we perform multiple quantile regression using the sum of pinball losses as strictly consistent
scoring functions. Since quantiles are monotone in their probability levels, we modify the FN network architecture
of Richman \cite{Richman3} such that we can guarantee this monotonicity. 

In a second step, we extend the work of Guillen et al.~\cite{Guillen2021} in several ways.
In risk management, besides knowing quantiles, one is also interested in estimating ES,
which reflects the average outcome beyond the quantile. 
Since ES is not elicitable \cite{Gneiting}, stand-alone ES regression is generally not possible.
Fissler--Ziegel \cite{FisslerZiegel2016} showed that the pair of ES together with the quantile at the same probability level is elicitable.
This paved the way to joint quantile--ES regression, see Dimitriadis--Bayer \cite{DimitriadisBayer} and Guillen et al.~\cite{Guillen2021} for (G)LM approaches.
We first extend the result of Fissler--Ziegel \cite{FisslerZiegel2016}, showing the elicitability of the \emph{composite triplet} consisting of a quantile together with the lower and upper ES at the same probability level, and characterizing the entire class of strictly consistent scoring functions for it.
Hence, we are in the position to estimate a full composite regression model in a one-step procedure.
This is in contrast to the two-step estimation approach of Barendse \cite{Barendse2020}, first estimating the quantile and then in a second step the ES below and above the quantile.
Such a two-step estimation approach becomes problematic in the presence of common parameters in the three components of the regression model.
Second, we motivate the particular choice of the strictly consistent scoring function in a data-driven manner, striving for efficient parameter estimation. 
To this end, we use optimality results known for mean regression which are akin to the optimality results derived in Dimitriadis et al.~\cite{DimitriadisFisslerZiegel2020} for the pair of the quantile and ES.
The third novelty we provide is the FN network architecture that
respects the necessary monotonicity property of the composite triplet in an estimation context.
This architecture can directly be fitted
using the stochastic gradient descent (SGD) algorithm. This fitting approach does neither require a two-step fitting approach nor the EM algorithm as it is used 
in related problems.
Moreover, our fitting turns out to be robust, and we do not encounter the
stability issues as in the two-step approaches and the EM algorithm.

\medskip

{\bf Organization of this manuscript.}
In Section \ref{section Statistical learning} we review the concepts of strictly consistent scoring functions used in estimation and forecast evaluation.
We recall known characterization results of strictly consistent scores for the mean and for the quantile.
We extend the results of Fissler--Ziegel \cite{FisslerZiegel2016} introducing the class of strictly consistent scoring functions for
the composite triplet of the quantile, lower ES and upper ES at the same probability level.
In Section \ref{Deep quantile and deep composite model regressions} we discuss how these
quantities can be estimated within a neural network regression framework, and we also discuss
asymptotically efficient choices of scoring functions.
In Section \ref{Claim size example} we give a real data example which demonstrates the suitability
of our proposal. Section \ref{Conclusions} concludes and provides an outlook to related open problems.


\section{Statistical learning}
\label{section Statistical learning}
\subsection{A review of strictly consistent scoring functions}
\label{Strictly consistent scoring and pinball loss for quantiles}
We start from the decision-theoretic approach of forecast evaluation developed by
Gneiting \cite{Gneiting} and Gneiting--Raftery \cite{GneitingRaftery}, also
used for M-estimation in Dimitriadis et al.~\cite{DimitriadisFisslerZiegel2020}.
This provides us with
suitable choices of scoring functions for model fitting, model selection, and forecast evaluation.
Denote by ${\cal F}$ the class of considered distribution functions $F \in {\cal F}$. 
Let $\Y \subseteq \R$ be an interval, a halfline or the whole real line $\R$, such that the support of each $F\in {\cal F}$ is contained in $\Y$. Choose an action space $\A \subseteq \R$ 
from which we select actions $a \in \A$ to 
estimate a statistics $A(F) \in \A$ of $F \in {\cal F}$. That is, we have a functional
\begin{equation}\label{law invariant statistics}
A :{\cal F} \to \A \qquad F \mapsto A(F),
\end{equation}
that we try to estimate. Commonly used functionals are the mean functional
$A(F)=\E_F[Y]$ for $Y\sim F \in {\cal F}$ and quantiles. 
Given a probability level $\tau \in (0,1)$,  the $\tau$-quantile of $F \in {\cal F}$
is given by the (left-continuous) generalized inverse
\begin{equation}\label{definition of generalized inverse}
A(F)=F^{-1}(\tau) = \inf \left\{ y \in \R; ~F(y) \ge \tau \right\}.
\end{equation}

To receive an intuitive understanding we usually speak about a functional \eqref{law invariant statistics}
that attains a single value $A(F)$ in the action space $\A$. Often this functional is obtained by an optimization
(M-estimator),
or by finding the roots of a given function (Z-estimator). 
Since, in general, we 
do not want to assume uniqueness of such a solution, we should understand this functional
as a set-valued map
\begin{equation}\label{law invariant statistics set valued}
A :{\cal F} \to {\cal P}({\A}) \qquad F \mapsto A(F) \subset \A,
\end{equation}
where ${\cal P}({\A})$ is the power set of $\A$. 
E.g., the set-valued $\tau$-quantile of a distribution $F$ is given by
\begin{equation}
\label{eq:quantile}
 q_\tau(F) = \{t\in\R\colon F(t-)\le \tau \le F(t)\},
\end{equation}
where $F(t-) = \lim_{x\uparrow t}F(x)$. This defines a closed interval and its lower endpoint corresponds to the 
left-continuous generalized inverse $F^{-1}(\tau)$ given in \eqref{definition of generalized inverse}.
To keep notation simple, we identify 
$\widehat{a}$ and $A(F)$ if $A(F)=\{\widehat{a}\}$ is a singleton.

In order to evaluate the accuracy of actions $a$ for the statistics $A(F)$ (for unknown $F$) we consider a scoring function (also called loss function)
\begin{equation*}
  L: \Y \times \A \to \R, \qquad (y; a)\mapsto L(y; a).
\end{equation*}
This describes the loss of an action $a \in \A$ if a realization  $y $ of $Y \sim F$ materializes.
To incentivize truthful forecasts,
Gneiting \cite{Gneiting} advocates that
this scoring function $L$ should be strictly consistent for the functional $F\mapsto A(F)$ of interest.

\begin{defi}[strict consistency] 
A scoring function $L: \Y \times \A \to \R$ is ${\cal F}$-consistent for a given functional
$A: {\cal F} \to {\cal P}(\A)$ if $\E_F \left[| L(Y;a)| \right]<\infty$ for all $Y \sim F \in {\cal F}$ and for all $a\in\A$ and
if
\begin{equation}
\label{consistency of prediction}
\E_F \left[ L(Y;\widehat{a}) \right] ~\le  ~ \E_F \left[ L(Y;a) \right],
\end{equation}
for all $Y \sim F \in {\cal F}$, $\widehat{a} \in A(F)$ and $a \in \A$.
It is strictly ${\cal F}$-consistent if it is ${\cal F}$-consistent and equality in \eqref{consistency of prediction}
implies that $a \in A(F)$.
\end{defi}
Gneiting \cite{Gneiting} shows that strictly consistent scoring functions are linked to proper scoring rules, which serve to evaluate probabilistic forecasts (i.e.~actions $a$ taking the form of probability distributions) and which are discussed in detail in 
Gneiting--Raftery \cite{GneitingRaftery}.
On the estimation side, we also need (strict) consistency of scoring functions to obtain consistency
of M-estimators, see Chapter 5 in 
Van der Vaart \cite{VanderVaart2} and Dimitriadis et al.~\cite{DimitriadisFisslerZiegel2020}.
This raises the question of which functionals $A: {\cal F} \to \A$ admit strictly consistent scoring functions $L$, i.e., are elicitable.

\begin{defi}[elicitable]
\label{definition elicitable}
The functional $A: {\cal F} \to {\cal P}(\A)$ is elicitable on a given class
of distribution functions ${\cal F}$ if there exists a scoring function $L$ that is strictly ${\cal F}$-consistent
for $A$.
\end{defi}

The general question of elicitability is studied and discussed in detail in Gneiting \cite{Gneiting}. For instance,
the mean functional $A(F)=\E_F[Y]$ and the $\tau$-quantile $q_\tau$ defined in \eqref{eq:quantile} are elicitable; see Theorems \ref{theorem Savage} and \ref{theorem Thomson}, below.
We start with useful properties of scoring functions $L$ that are going to be used in the
next theorems. Assume $\Y=\A \subseteq \R$
is an interval with non-empty interior. We set:
\begin{itemize}
\item[(S0)] $L(y;a) \ge 0$ and equality holds if and only if $y=a$;
\item[(S1)] $L(y;a)$ is measurable in $y$ and continuous in $a$;
\item[(S2)] The partial derivative $\partial_a L(y;a)$ exists and is continuous in $a$ whenever $a\neq y$.
\end{itemize}
These regularity conditions can often be weakened. Especially the positivity in (S0) can be relaxed. 
However, it is a nice property to have
in optimizations as it gives us a natural lower bound to the minimal score that can be achieved.

The following result goes back to Savage \cite{Savage}.
\begin{theo}[Gneiting \cite{Gneiting}, Theorem 7]
\label{theorem Savage}
Let ${\cal F}$ be a class of distribution functions on interval  $\Y \subseteq \R$ with finite first moment.
\begin{itemize}
\item The mean functional  $F\mapsto A(F)=\E_F[Y]$ is elicitable relative to ${\cal F}$.
\item Assume the scoring function $L:\Y\times \A \to \R_+$ satisfies (S0)--(S2) for an interval $\Y=\A \subseteq \R$
and that ${\cal F}$ is the class 
of compactly supported distributions on $\Y$.
The scoring function
$L$ is ${\cal F}$-consistent for the mean functional if and only if $L$ is of the
form
\begin{equation}\label{Bregman divergence definition}
L(y;a) =L_{\phi}(y;a) = \phi(y) - \phi(a)-\phi'(a) \left(y-a\right),
\end{equation}
for a convex function $\phi$ with (sub-)gradient $\phi'$ on $\Y$.
\item If $\phi$ is strictly convex on $\Y$, then scoring function 
\eqref{Bregman divergence definition} is strictly consistent for the mean functional on the class
${\cal F}$ of distribution functions $F$ on $\Y$ for which both $\E_F[Y]$
and $\E_F[\phi(Y)]$ exist and are finite.
\end{itemize}
\end{theo}

The maps $L_{\phi}$ defined in \eqref{Bregman divergence definition} are called {\it Bregman
divergences} and, basically, the previous theorem says that all strictly consistent scoring functions
for the mean functional are Bregman divergences.
The most prominent Bregman divergence arises by setting $\phi(y) = y^2$, yielding the squared loss $L(y;a) = (y-a)^2$.
On the other
hand, the variance functional $\mathbb V_F(Y)$ is not elicitable 
(Gneiting \cite{Gneiting} and Osband \cite{Osband1985}), i.e., there does not exist any strictly consistent
scoring function for estimating the variance according to a minimization \eqref{consistency of prediction}.

The following elicitability results for quantiles \eqref{eq:quantile} originates from Thomson \cite{Thomson} and Saerens \cite{Saerens}.
\begin{theo}[Gneiting \cite{Gneiting}, Theorem 9]
\label{theorem Thomson}
Let ${\cal F}$ be a class of distribution functions on interval  $\Y \subseteq \R$  and choose $\tau \in (0,1)$.
\begin{itemize}
\item The $\tau$-quantile \eqref{eq:quantile} is elicitable relative to ${\cal F}$.
\item Assume the scoring function $L:\Y\times \A \to \R_+$ satisfies (S0)--(S2) for an interval $\Y=\A \subseteq \R$
and that ${\cal F}$ is the class 
of compactly supported distributions on $\Y$.
The scoring function
$L$ is ${\cal F}$-consistent for the $\tau$-quantile \eqref{eq:quantile} if and only if $L$ is of the
form
\begin{equation}
\label{definition pinball loss}
L_\tau(y;a) = \left(g(y) -g(a) \right)\left(\tau-\mathds{1}_{\{y \le a\}} \right),
\end{equation}
for a non-decreasing function $g$ on $\Y$.
\item If $g$ is strictly increasing on $\Y$ and $\E_F[g(Y)]$ exists and is finite for all $F\in\mathcal F$, then  $L$ defined by \eqref{definition pinball loss}
is strictly ${\cal F}$-consistent for the $\tau$-quantile \eqref{eq:quantile}.
\end{itemize}
\end{theo}
Members of the class \eqref{definition pinball loss} are called \emph{generalized piecewise linear losses}, see Gneiting \cite{Gneiting2011b}.
Basically, it is this theorem that allows us to consider quantile regression, introduced
by Koenker--Bassett \cite{KoenkerBassett}, as it tells us that quantiles can be estimated
from strictly consistent scoring functions. This is going to be outlined below.
There is still the freedom of the choice of the strictly increasing function $g$. The most commonly
used choice is the identity function $g(y)=y$. This then provides us with the so-called
{\it pinball loss} for given probability level $\tau \in (0,1)$
\begin{equation}
\label{pinball loss}
L_\tau(y;a) = (y-a)\left(\tau - \mathds{1}_{\{ y \le a \}}\right)~\ge ~0.
\end{equation} 
In view of Theorem \ref{theorem Thomson}, 
the choice of pinball loss  requires that ${\cal F}$ contains only distributions  with a finite first moment.

We introduce scoring functions closely related to the pinball loss \eqref{pinball loss} where the positivity assumption (S0) has been relaxed:
\begin{align*}
S^-_\tau(y;a) &= \left(\mathds{1}_{\{y\le a\}} - \tau\right)a - \mathds{1}_{\{y\le a\}}y, \\
S^+_\tau(y;a) & = \left(1-\tau - \mathds{1}_{\{y > a\}}\right)a + \mathds{1}_{\{y> a\}}y,
\end{align*} for $y,a\in\R$ and for $\tau \in (0,1)$.
Note that $S^+_\tau(y;a) = S^-_\tau(y;a) + y$, moreover, it holds for the pinball loss
\begin{equation*}
L_\tau(y;a) = S^-_\tau(y;a) +\tau y= S^+_\tau(y;a)-(1-\tau)y.
\end{equation*} 
An immediate consequence from Theorem \ref{theorem Thomson} is the
following corollary.
\begin{cor}\label{cor:ES}
Let $\mathcal F$ contain only distributions with finite first moments.
Then $S^-_\tau$ and $S^+_\tau$ are strictly $\mathcal F$-consistent for the $\tau$-quantile  \eqref{eq:quantile},
that is, 
\begin{equation}
\label{eq:argmin}
q_\tau(F)=\argmin_{a\in\R} \mathbb E_F\big[ S^-_\tau(Y;a) ] = \argmin_{a\in\R} \mathbb E_F\big[ S^+_\tau(Y;a) ]
= \argmin_{a\in\R} \mathbb E_F\big[ L_\tau(Y;a) ].
\end{equation}
\end{cor}
Observe that the three functions $S^-_\tau$
$S^+_\tau$ and $L_\tau$ only differ in terms that do not depend on $a$, and therefore the
argument of these minimizations is the same. The pinball loss $L_\tau(y;a)$ has 
the advantage to satisfy condition (S0).

\subsection{Expected shortfall and the composite triplet}
From the previous section we know that we can estimate $\tau$-quantiles
\eqref{eq:quantile}
 by minimizing
\eqref{consistency of prediction} under the strictly consistent pinball loss \eqref{pinball loss}.
In actuarial science we are often interested in also considering the lower ES and the upper ES
\begin{equation}\label{expectation shortfall}
{\ES}^-_{\tau}(Y)=  \frac{1}{\tau}\int_0^\tau F^{-1}(p)\,\mathrm{d}p
\qquad \text{ and } \qquad
{\ES}^+_{\tau}(Y)=  \frac{1}{1-\tau}\int_\tau^{1}F^{-1}(p)\,\mathrm{d}p,
\end{equation}
of a random variable $Y \sim F$. 
Since $\ES_\tau^-$ and $\ES_\tau^+$ are law-determined, we can interpret them as functionals \eqref{law invariant statistics}.
The monotonicity of the generalized inverse $p\mapsto F^{-1}(p)$ immediately yields
\begin{equation}
\label{eq:monotone}
\ES_\tau^-(F)\le F^{-1}(\tau) \le \ES_\tau^+(F).
\end{equation}

\begin{rem} \normalfont
If the distribution function $F$ is continuous, in particular, if $F(F^{-1}(\tau)) = \tau$, 
the ES is equal to the more familiar conditional tail
expectation (CTE), see Lemma 2.16 in McNeil et al.~\cite{McNeil}. Namely, we have
\begin{equation*}
{\ES}^-_{\tau}(Y)=
{\rm CTE}^-_\tau(Y)=\E_F \left[ Y \left| Y \le F^{-1}(\tau) \right] \right. ,
\end{equation*}
and
\begin{equation}\label{conditional tail expectation}
{\ES}^+_{\tau}(Y)=
{\rm CTE}^+_\tau(Y)=\E_F \left[ Y \left| Y > F^{-1}(\tau) \right] \right.. 
\end{equation}
If we can estimate these two CTEs, this allows us to fit a composite model to $Y$, with the quantile $F^{-1}(\tau)$
giving the splicing point where lower and  upper parts are concatenated.
\end{rem}

The ES, considered as functionals, turn out not to be elicitable on families of distributions $\mathcal F$ which offer the necessary flexibility in most modeling situations; see Gneiting \cite{Gneiting} and Weber \cite{Weber2006}. Thus, there is no strictly consistent scoring function suitable for M-estimation of
ES. 
Fissler--Ziegel \cite{FisslerZiegel2016} have proved that
${\rm ES}^-_\tau(Y)$ is {\it jointly} elicitable with the $\tau$-quantile $F^{-1}(\tau)$, the latter also being called Value-at-Risk (VaR)
in risk management. We are going to extend this result in Theorem \ref{theo joint elicitability 2}, establishing the elicitability of the \emph{composite triplet}
$({\ES}^-_\tau, q_\tau, {\ES}^+_\tau)$.
We first need the following lemma.

\begin{lemma}\label{lem:ES}
For any distribution $F$ with finite first moment, it holds that 
\begin{equation}
\label{eq:min}
\ES^-_\tau(F) = -\frac{1}{\tau} \min_{v\in\R} \mathbb E_F\big[ S^-_\tau(Y;v) ]
\qquad \text{ and } \qquad
\ES^+_\tau(F) = \frac{1}{1-\tau}\min_{v\in\R} \mathbb E_F\big[ S^+_\tau(Y;v) ].
\end{equation}
\end{lemma}

\medskip

\textit{Proof.}
 This follows directly along the lines of Lemmas 2.3 and 3.3 in
 Embrechts--Wang \cite{EmbrechtsWang2015}.
\EndProof

\medskip

\begin{theo}
\label{theo joint elicitability 2}
Choose $\tau\in(0,1)$ and let $\mathcal F$ contain only distributions with finite first moments and supported on $\Y\subseteq \R$.
The scoring function
$L\colon \Y\times\Y^3 \to \R_+$ of the form
\begin{eqnarray}
\label{eq:score}
L(y; e^-,v,e^+)&=&\left(g(y) -g(v) \right)\left(\tau-\mathds{1}_{\{y \le v\}} \right)
\\ \nonumber& 
& + ~\left\langle \nabla \Phi(e^-, e^+), \begin{pmatrix}
e^- + \tfrac{1}{\tau}S^-_\tau(y;v) \\
e^+ - \tfrac{1}{1-\tau}S^+_{\tau}(y;v)
\end{pmatrix}
\right\rangle -\Phi(e^-, e^+)+ \Phi(y,y), 
\end{eqnarray}
is (strictly) $\mathcal F$-consistent for the composite triplet $({\ES}^-_\tau, q_\tau, {\ES}^+_\tau)$ if 
$\Phi\colon \Y^2\to\R$ is (strictly) convex with sub-gradient $\nabla \Phi$ such that for $g\colon \Y \to\R$ and for all $(e^-,e^+)\in \Y^2$ the function
\begin{equation}
\label{eq:increasing}
G_{e^-, e^+}\colon \Y \to \R, 
\quad v\mapsto g(v) + \tfrac{1}{\tau} \partial_1\Phi(e^-,e^+)v - \tfrac{1}{1-\tau} \partial_2 \Phi(e^-,e^+)v
\end{equation}
is (strictly) increasing,
and if $\E_F[|g(Y)|]<\infty$, $\E_F[|\Phi(Y,Y)|]<\infty$ for all $Y\sim F\in \mathcal F$.
Moreover, $L(y; e^-,v,e^+) \ge L(y; y,y,y) = 0$.
\end{theo}

The intuitive interpretation is that the score in \eqref{eq:score} is a combination of a Bregman divergence (second line) and a generalized piecewise linear loss (first line in combination with $S_{\tau}^-$ and $S_{\tau}^+$).
This structure will be exploited in the proof.

\medskip

\textit{Proof of Theorem \ref{theo joint elicitability 2}.}
First, fix some $v\in\Y$. The map
\[
(y;e^-,e^+)\mapsto L(y;e^-,v,e^+) = \left\langle \nabla \Phi(e^-, e^+), \begin{pmatrix}
e^- + \tfrac{1}{\tau}S^-_\tau(y;v) \\
e^+ - \tfrac{1}{1-\tau}S^+_{\tau}(y;v)
\end{pmatrix}
\right\rangle -\Phi(e^-, e^+) + b_{v}(y),
\]
where the remainder $b_v(y)$ does not depend on $(e^-, e^+)$,
is a Bregman divergence. Hence, if $\Phi$ is (strictly) convex, it is (strictly) $\F$-consistent for the functional
\[
F\mapsto \big(-\tfrac{1}{\tau} \mathbb E_F\big[ S^-_\tau(Y;v) ], \tfrac{1}{1-\tau} \mathbb E_F\big[ S^+_\tau(Y;v) ]\big).
\]
Second, for fixed $(e^-,e^+)\in \Y^2$, the map
\[
(y;v)\mapsto L(y;e^-,v,e^+) = (\one_{\{y\le v\}} - \tau)G_{e^-, e^+}(v) - \one_{\{y\le v\}} G_{e^-, e^+}(y) + b_{e^-,e^+}(y),
\]
where the remainder $b_{e^-,e^+}(y)$ does not depend on $v$, 
is a generalized piecewise linear loss \eqref{definition pinball loss} not necessarily satisfying the positivity (S0). 
Hence, if $G_{e^-, e^+}$ is (strictly) increasing, it is (strictly) $\F$-consistent for $q_\tau$.

Combining these two observations yields the (strict) $\F$-consistency of $L$. 

Finally, $L(y;y,y,y)=0$ can be verified by a direct computation, and the non-negativity of $L$ follows from its consistency and the fact that for a random variable $Y$ which deterministically equals the constant $y\in\R$, it holds that $\ES_\tau^-(Y) = \ES_\tau^+(Y) = y$ and $q_\tau(Y) = \{y\}$.
\EndProof

\medskip

Theorem \ref{theo joint elicitability 2} extends the result by Frongillo--Kash \cite[Theorem 1]{FrongilloKash2020} asserting that an elicitable functional (in our case the $\tau$-quantile) is jointly elicitable with finitely many associated Bayes risks (here corresponding to $\ES_\tau^-$ and $\ES_\tau^+$).

\medskip

Theorem \ref{thm:necessary}, below, shows that the scores of the form \eqref{eq:score} are basically the only $\F$-consistent scores for $(\ES_\tau^-, q_\tau, \ES_\tau^+)$. 
The argument -- which can be found in its entirety in the Supplementary Section \ref{Suppl:Characterisation} -- uses Osband's principle, which originates from Kent Osband's seminal thesis \cite{Osband1985}; see Gneiting \cite{Gneiting} for an intuitive exposition and Fissler--Ziegel \cite{FisslerZiegel2016} for a precise technical formulation.
It  exploits first- and second-order conditions stemming from the optimization problem of (strict) consistency \eqref{consistency of prediction}. Hence, we need to impose smoothness conditions on the expected score, which play the role of condition (S2), but are slightly weaker. This is reflected in Assumption \ref{ass:S2}, and the fact that we work within a subclass of $\F\subseteq \F_{\rm cont}^\tau$, the class of distribution functions on $\R$ which are continuously differentiable and whose $\tau$-quantiles are singletons.
The second kind of conditions are richness conditions on the underlying class of distributions $\F$. On the one hand, the first- and second-order conditions only yield local assertions about the expected scores. Hence, $\F$ needs to be rich enough such that the functional maps surjectively to the action domain considered.
Recalling monotonicity condition \eqref{eq:monotone}, this means that we can only provide conditions on action spaces contained in $\{(a_1,a_2,a_3)\in\R^3\colon a_1\le a_2\le a_3\}$.
Second, the richness conditions ensure that the functional `varies sufficiently' such that it can be distinguished from any other functional.\footnote{E.g., on the class of symmetric distributions with positive densities, the mean and the median coincide and cannot be distinguished. Such phenomena need to be excluded.}
This is reflected in Assumptions \ref{ass:V1} and \ref{ass:V4}.
Finally, since Osband's principle actually characterizes the \emph{gradient} of the \emph{expected} score, one needs to be in the position to integrate this gradient (Assumption \ref{ass:VS1}) and to approximate the pointwise values of the score with expectations (Assumption \ref{ass:F1}).

\begin{theo}\label{thm:necessary}
Let $\tau\in(0,1)$ and $\F\subseteq \F_{\rm cont}^\tau$.
Let $L\colon \R\times \A\to\R$, $\A\subseteq \{(a_1,a_2,a_3)\in\R^3\colon a_1\le a_2\le a_3\}$ be an $\F$-consistent scoring function for the composite triplet $({\ES}^-_\tau, q_\tau, {\ES}^+_\tau)$, satisfying Assumptions \ref{ass:V1}--\ref{ass:VS1} and $L(y;y,y,y)=0$ for all $y\in\R$ such that $(y,y,y)\in\A$.
Then $L$ is necessarily of the form \eqref{eq:score} almost everywhere where $\Phi$ is convex and 
where for any fixed $(e^-,e^+)\in\R^2$ such that there is a $v\in\R$ with $(e^-,v,e^+)\in\A$ the function $G_{e^-,e^+}$ in \eqref{eq:increasing} is increasing.
\end{theo}

Supplementary Section \ref{Suppl:Characterisation} provides all technical details. In particular, the proof of Theorem \ref{thm:necessary} can be found in Subsection \ref{subsec:proof}.

\subsection{Particular choices of scoring functions for the composite triplet}
\label{subsec:choice}

There remain the choices of the functions $g$ and $\Phi$ in \eqref{eq:score}. Especially, the choice of the strictly convex function $\Phi$ such that $G_{e^-, e^+}$ in \eqref{eq:increasing} is strictly increasing  is not obvious. 
We discuss the following particularly convenient three choices for given $\tau \in (0,1)$
\begin{align}
\label{eq:Phi1}
\Phi(e^-,e^+) &= \phi_-(e^-) + \phi_+(e^+), \\
\label{eq:Phi2}
\Phi(e^-,e^+) &= \phi(\tau e^- + (1-\tau) e^+) + \phi_+(e^+),\\
\label{eq:Phi3}
\Phi(e^-,e^+) &= \phi(\tau e^- + (1-\tau) e^+) + \phi_-(e^-),
\end{align}
where $\phi, \phi_+, \phi_-\colon \Y \to \R$ are strictly convex and the (sub-)gradients satisfy $\phi'_+<0$, $\phi'_->0$.
Moreover, if we choose $g\colon \Y \to \R$ to be an increasing function (not necessarily strictly increasing), we obtain for $G_{e^-, e^+}$, defined in \eqref{eq:increasing},
\begin{align*}
G_{e^-, e^+}(v) 
= \begin{cases}
g(v) + \tfrac{1}{\tau} \phi'_-(e^-)v - \tfrac{1}{1-\tau}\phi'_+(e^+)v \qquad & \text{for $\Phi$ in \eqref{eq:Phi1}}, \\
g(v) - \tfrac{1}{1-\tau}\phi'_+(e^+)v & \text{for $\Phi$ in \eqref{eq:Phi2}}, \\
g(v) + \tfrac{1}{\tau} \phi'_-(e^-)v & \text{for $\Phi$ in \eqref{eq:Phi3}}. \\
\end{cases}
\end{align*}
Since $\phi'_+<0$ and $\phi'_->0$, $G_{e^-, e^+}$ is strictly increasing, even if $g$ is constant.
This yields the following three types of scores: 
For $\Phi$ in \eqref{eq:Phi1} we get
\begin{eqnarray}
\label{eq:score additive decomp}
L(y; e^-,v,e^+)&=&\left(g(y) -g(v) \right)\left(\tau-\mathds{1}_{\{y \le v\}} \right)
\\ \nonumber& 
& + ~\phi'_-(e^-) \left(e^- + \tfrac{1}{\tau}S^-_{\tau}(y;v) \right) - \phi_-(e^-) + \phi_-(y)
\\\nonumber& & 
+ ~\phi'_+(e^+) \left(e^+ - \tfrac{1}{1-\tau}S^+_{\tau}(y;v) \right) - \phi_+(e^+) + \phi_+(y).
\end{eqnarray}
Similarly, for $\Phi$ in \eqref{eq:Phi2} we receive
\begin{eqnarray}
\label{eq:score revelation}
L(y; e^-,v,e^+)&=&\left(g(y) -g(v) \right)\left(\tau-\mathds{1}_{\{y \le v\}} \right)
\\ \nonumber& 
& + ~\phi'_+(e^+) \left(e^+ - \tfrac{1}{1-\tau}S^+_{\tau}(y;v) \right) - \phi_+(e^+) + \phi_+(y)
\\\nonumber& & 
+ ~\phi'(\tau e^- + (1-\tau)e^+) \left(\tau e^- + (1-\tau)e^+ -y \right) - \phi(\tau e^- + (1-\tau)e^+) + \phi(y),
\end{eqnarray}
and finally for \eqref{eq:Phi3}
\begin{eqnarray}
\label{eq:score revelation2}
L(y; e^-,v,e^+)&=&\left(g(y) -g(v) \right)\left(\tau-\mathds{1}_{\{y \le v\}} \right)
\\ \nonumber& 
& + ~\phi'_-(e^-) \left(e^- + \tfrac{1}{\tau}S^-_{\tau}(y;v) \right) - \phi_-(e^-) + \phi_-(y)
\\\nonumber& &  
+ ~\phi'(\tau e^- + (1-\tau)e^+) \left(\tau e^- + (1-\tau)e^+ -y \right) - \phi(\tau e^- + (1-\tau)e^+) + \phi(y).
\end{eqnarray}
Scores of the form \eqref{eq:score additive decomp} can directly be interpreted to be the sum of scores for the pair $(q_\tau, \ES_\tau^-)$ as deduced in Fissler--Ziegel \cite{FisslerZiegel2016} and for the pair $(q_\tau, \ES_\tau^+)$ as derived in Nolde--Ziegel \cite{NoldeZiegel2017}.
On the other hand, \eqref{eq:score revelation} can be deduced from the sum of a scoring function for $(q_\tau, \ES_\tau^+)$ and for the mean, using the so-called \emph{revelation principle}. This principle originates from Osband's thesis \cite{Osband1985} and it has been made rigorous in Gneiting \cite[Theorem 4]{Gneiting}. It asserts that any bijection of an elicitable functional is elicitable and it makes the corresponding strictly consistent scoring functions explicit.
In the case of \eqref{eq:score additive decomp}, this bijection is 
\[
(\ES_\tau^-,q_\tau, \ES_\tau^+) \mapsto (q_\tau, \ES_\tau^+, \tau \ES_\tau^- + (1-\tau)\ES_\tau^+) = (q_\tau, \ES_\tau^+, \E).
\]
For the scores in \eqref{eq:score revelation2}, the corresponding bijection reads similar.

\medskip

The next section shows how to use strictly consistent scoring functions for parameter estimation via M-estimation in a regression context.
Here, the (strict) $\F$-consistency of the corresponding score is crucial to obtain the consistency of the M-estimator, i.e., that the M-estimator converges in probability to the true value as the sample size goes to infinity; see Dimitriades et al.~\cite{DimitriadisFisslerZiegel2020}.
Knowing that the estimator is consistent, it is of interest to have fast convergence, i.e., an efficient estimator.
Under asymptotic normality of the estimator, a more efficient estimator has a strictly smaller asymptotic covariance matrix.\footnote{As usual, we mean this with respect to the Loewner order. That is, a covariance matrix $A$ is strictly smaller than a covariance matrix $B$ of the same dimension if $A\neq B$ and if $B-A$ is positive semi-definit.} 
In the absence of any explanatory information (that is, when estimating an intercept only model), the choice of the strictly consistent scoring function is immaterial since the intercept estimators under different
strictly consistent scores will always coincide on finite samples and correspond to the functional of the empirical distribution function.
In a more interesting regression scenario, however, using feature information, the question of efficiency becomes important,
this is discussed in our setup in Section \ref{Deep composite model regression}, below.


\section{Deep quantile and deep composite model regressions}
\label{Deep quantile and deep composite model regressions}
\subsection{Quantile regression}

The previous sections have discussed estimation theory of functionals
of (unkown) distribution functions $F$. We now lift this framework to a regression context where random variables
$Y$ are supported by covariates (feature information).
Assume that the random variable $Y$ is established with feature information $\bX \in {\cal X} \subset \{1\}\times \R^q$,
where ${\cal X}$ is the feature space of all potential explanatory variables $\bX$. We assume for a datum
$(Y,\bX)$ that the conditional distribution of $Y$, given $\bX$, is described by a distribution
function $F_{Y|\bX = \bx}$, $\bx\in{\cal X}$, or in short $F_{Y|\bx}$, and the conditional $\tau$-quantile of this random variable is given by
the left-continuous generalized inverse
\begin{equation*}
F_{Y|\bx}^{-1}(\tau) = \inf \left\{ y \in \R; ~F_{Y|\bx}(y) \ge \tau \right\}.
\end{equation*}
Note that we label distributions with subscripts $Y|\bx$, now, to clearly indicate that we
are considering the conditional distribution of $Y$, given feature information $\bX = \bx \in {\cal X}$.
This gives us the existence of a regression function $Q_\tau :{\cal X} \to \R$ 
such that
\begin{equation*}
\bx ~ \mapsto ~ Q_\tau(\bx)=F_{Y|\bx}^{-1}(\tau).
\end{equation*}
Quantile regression tries to determine this regression function $Q_\tau$ from a given function class ${\cal Q}$ based
on i.i.d.~observations $(Y_i,\bX_i)$, $1\le i \le n$. The classical approach
of Koenker--Bassett \cite{KoenkerBassett} makes a GLM
assumption by postulating the existence of a  strictly monotone and smooth link function $h:\R \to \R$
such that 
\begin{equation}\label{quantile GLMs}
Q_\tau(\bx) = h^{-1} \langle \bbeta, \bx \rangle,
\end{equation}
with regression parameter $\bbeta \in \R^{q+1}$ and $\langle \cdot, \cdot \rangle$ denoting the
scalar product in the Euclidean space $\R^{q+1}$. In this case the function class ${\cal Q}$
is parametrized by a regression parameter $\bbeta \in \R^{q+1}$, that we try to optimally determine
from i.i.d.~data $(Y_i,\bX_i)$, $1\le i \le n$. The choice of the link $h$ acts as a hyper-parameter, 
that is not part of the optimization process.

We can choose a strictly consistent scoring function $L_\tau$ \eqref{definition pinball loss} for the $\tau$-quantile $F_{Y|\bx}^{-1}(\tau)$ and estimate the regression parameter
$\bbeta \in \R^{q+1}$ by
\begin{equation*}
\widehat{\bbeta}_\tau ~= ~\underset{\bbeta \in \R^{q+1}}{\arg\min}~
\E \left[ L_\tau\left(Y;  h^{-1} \langle \bbeta, \bX \rangle\right) \right],
\end{equation*}
subject to existence. Typically, we do not know the true distribution function and, therefore,
cannot explicitly evaluate the right-hand side of the above optimization. Using an empirical version based on
i.i.d.~data $(Y_i,\bX_i)$, $1\le i \le n$, motivates M-estimator
\begin{equation}\label{MLE GLM}
\widehat{\bbeta}_\tau ~= ~\underset{\bbeta \in \R^{q+1}}{\arg\min}~
\frac{1}{n} \sum_{i=1}^n L_\tau\left(Y_i;  h^{-1} \langle \bbeta, \bX_i \rangle\right).
\end{equation}

\subsection{Deep quantile regression}
\label{subsec:Deep quantile regression}
In practice, the GLM structure \eqref{quantile GLMs} is often too restrictive.
This emphasizes to use a more flexible function class ${\cal Q}$. FN networks
provide the building blocks for such a more flexible function class. 
This motivates {\it deep quantile regression} which has been 
introduced to the actuarial literature by Richman \cite{Richman3}.
We replace 
\eqref{quantile GLMs} by
\begin{equation}\label{quantile NNs}
Q_\tau(\bx) = h^{-1} \langle \bbeta, \bz^{(d:1)}(\bx) \rangle,
\end{equation}
where $\bz^{(d:1)}: {\cal X} \to \{1\}\times \R^{r_d}$ is an FN network of depth $d \in \N$,  regression parameter 
$\bbeta \in \R^{r_d+1}$ and link $h$.
The FN network $\bz^{(d:1)}$ is a composition of $d$ FN layers
\begin{equation*}
\bx \in {\cal X} ~\mapsto ~\bz^{(d:1)}(\bx)=\left(\bz^{(d)}\circ \cdots \circ \bz^{(1)}\right)(\bx) ~\in~ \{1\}\times \R^{r_d},
\end{equation*}
with FN layers $\bz^{(m)}$ for $1\le m \le d$ involving further parameters
$\bw^{(m)}_j \in \R^{r_{m-1}+1}$, and with $r_{m-1}+1$ describing the input
dimension to FN layer $\bz^{(m)}$; for a detailed exhibition of FN networks we refer to
Section 7.2 in W\"uthrich--Merz \cite{WM2021}. Altogether this FN network
approach \eqref{quantile NNs} is parametrized by 
\begin{equation*}
\bvartheta =(\bw^{(1)}_1, \ldots, \bw^{(d)}_{r_d}, \bbeta)\in \R^r
\qquad \text{ of dimension }r= \sum_{m=1}^d r_m(r_{m-1}+1) + (r_d+1).
\end{equation*}
Thus, we fix a depth $d \in \N$, FN layer dimensions $r_1,\ldots, r_d \in \N$, the activation
functions in the FN layers and link function $h$ (as hyper-parameters), then the function class ${\cal Q}$
is parametrized by $\bvartheta$, and the ``optimal'' member  for i.i.d.~data $(Y_i,\bX_i)$, $1\le i \le n$, is found by
\begin{equation}\label{MLE NN}
\widehat{\bvartheta}_\tau ~= ~\underset{\bvartheta \in \R^{r}}{\arg\min}~
\frac{1}{n} \sum_{i=1}^n L_\tau\left(Y_i; h^{-1} \langle \bbeta, \bz^{(d:1)}(\bX_i) \rangle\right).
\end{equation}
On purpose ``optimal'' has been written in quotation marks. On a finite sample of size $n$
the solution to \eqref{MLE NN} will likely in-sample overfit to the learning data 
${\cal L}=(Y_i,\bX_i)_{1\le i \le n}$ because already small networks
are fairly flexible. Therefore, this model is usually fit with a SGD algorithm that explores an {\it early stopping rule}, i.e., that selects an estimate 
$\widehat{\bvartheta}_\tau$ that describes the systematic effects in the data ${\cal L}$
and not the noisy part.
This fitting approach is state-of-the-art, and it is described in detail in 
Section 7.2.3 of W\"uthrich--Merz \cite{WM2021}. Therefore, we will not repeat it here.

\subsection{Deep multiple quantile regression}
\label{Deep multiple quantile regression}

Often we do not want to estimate quantiles for only one probability level, but we would like to study
quantiles at different levels. For illustrative purposes we consider two probability levels $0<\tau_1 < \tau_2 < 1$,
and a generalization to more than two probability levels is straightforward. A naive way is
to individually estimate regression functions $\bx \mapsto Q_{\tau_l}(\bx)=F_{Y|\bx}^{-1}(\tau_l)$
 for $l=1,2$ using \eqref{quantile NNs}
and \eqref{MLE NN}. We call this a naive approach because these individual estimations
may violate the monotonicity property of quantiles, i.e., for all $\bx$ we require
\begin{equation}\label{multiple quantile case}
Q_{\tau_1}(\bx) ~ \le ~ Q_{\tau_2}(\bx).
\end{equation}
To simplify this outline, we assume that the random variable $Y$ is positive, a.s., which
implies that the generalized inverse $\tau \mapsto F_{Y|\bx}^{-1}(\tau) > 0$ has a positive
range. This motivates the choice of a link function $h$ with positive support $\R_+$.
For enforcing the monotonicity of quantiles for different levels \eqref{multiple quantile case}, we propose to jointly model these
quantiles. For our first proposal we choose two link functions $h$ and $h_+$
being both positively supported. In analogy to \eqref{quantile NNs}, this motivates  
{\it joint deep quantile regression} for probability levels
$\tau_1 <\tau_2$
\begin{eqnarray}\label{additive multiple quantile regression}
\bx & \mapsto& \left(Q_{\tau_1}(\bx),~Q_{\tau_2}(\bx)\right)^\top
\\&&=\left( h^{-1}\langle \bbeta_{\tau_1}, \bz^{(d:1)}(\bx) \rangle, ~
h^{-1}\langle \bbeta_{\tau_1}, \bz^{(d:1)}(\bx) \rangle+h_+^{-1}\langle \bbeta_{\tau_2}, \bz^{(d:1)}(\bx) \rangle \right)^\top ~\in ~\R_+^2,\nonumber
\end{eqnarray}
for a network parameter $\bvartheta=(\bw^{(1)}_1, \ldots, \bw^{(d)}_{r_d}, \bbeta_{\tau_1}, \bbeta_{\tau_2})^\top$.
Thus, we choose a common deep FN network $\bz^{(d:1)}$ that is {\it shared} by both quantiles, and the bigger
quantile $Q_{\tau_2}(\bx)$ is modeled by a positive difference
$h_+^{-1}\langle \bbeta_{\tau_2}, \bz^{(d:1)}(\bx)\rangle \ge 0$ to the smaller quantile $Q_{\tau_1}(\bx)$.
We call \eqref{additive multiple quantile regression} an {\it additive approach}  with base level 
$Q_{\tau_1}(\bx)$, and in network jargon we say that the FN network learns a common
representation $\bz_i = \bz^{(d:1)}(\bx_i)$ of features $\bx_i$, $1\le i \le n$, which is then used
in the two GLMs
\begin{equation*}
\bz_i ~\mapsto ~
\left(Q_{\tau_1}(\bz_i),~Q_{\tau_2}(\bz_i)\right)^\top
=\left( h^{-1}\langle \bbeta_{\tau_1}, \bz_i \rangle, ~
h^{-1}\langle \bbeta_{\tau_1}, \bz_i \rangle+h_+^{-1}\langle \bbeta_{\tau_2}, \bz_i \rangle \right)^\top ~\in ~\R_+^2.
\end{equation*}

Alternatively, for positive random variables $Y$,  a.s., we can choose the upper quantile $Q_{\tau_2}(\bx)$ as base level, and to ensure positivity we can multiplicatively
decrease this upper quantile. For this we choose the sigmoid function for $h_\sigma^{-1}=(1+\exp\{-x\})^{-1} \in (0,1)$
which motivates the {\it multiplicative approach}
for probability levels $\tau_1 <\tau_2$
\begin{eqnarray}\label{multiplicative multiple quantile regression}
\bx & \mapsto& \left(Q_{\tau_1}(\bx),~Q_{\tau_2}(\bx)\right)^\top
\\&&=\left( 
h_\sigma^{-1}\langle \bbeta_{\tau_1}, \bz^{(d:1)}(\bx) \rangle~
h^{-1}\langle \bbeta_{\tau_2}, \bz^{(d:1)}(\bx) \rangle, ~
h^{-1}\langle \bbeta_{\tau_2}, \bz^{(d:1)}(\bx) \rangle \right)^\top~
\in ~\R_+^2.\nonumber
\end{eqnarray}
Also in this case monotonicity is guaranteed because by assumption
$h_\sigma^{-1}\langle \bbeta_{\tau_1}, \bz^{(d:1)}(\bx) \rangle \in (0,1)$.

Since the learned representations $\bz_i = \bz^{(d:1)}(\bx_i)$ need to fit both quantiles simultaneously 
we need to learn these representations jointly. This motivates the optimization problem 
under regression assumption \eqref{additive multiple quantile regression}
or \eqref{multiplicative multiple quantile regression}, respectively, and up to over-fitting (see discussion after \eqref{MLE NN})
\begin{equation}\label{MLE NN multiple}
\widehat{\bvartheta}_{\tau_1,\tau_2} ~= ~\underset{\bvartheta }{\arg\min}~
\frac{1}{n} \sum_{i=1}^n \eta_1 L_{\tau_1}\left(Y_i; Q_{\tau_1}(\bX_i) \right)
+\eta_2 L_{\tau_2}\left(Y_i;  Q_{\tau_2}(\bX_i)\right),
\end{equation}
where the weights $\eta_1,\eta_2 >0$ are chosen such that both quantiles contribute roughly equally to the total score.
This is used to ensure that the quality of the estimate of both quantiles is roughly similar.

\subsection{Deep composite model regression}
\label{Deep composite model regression}
A deep composite regression model now only requires little changes compared to the deep multiple
quantile regression. Again, we assume that $Y>0$, a.s. We then aim at estimating the composite triplet
\begin{equation*}
{\rm ES}^-_\tau(Y|\bx) ~\le ~F_{Y|\bx}^{-1}(\tau) ~ \le ~ 
{\rm ES}^+_\tau(Y|\bx),
\end{equation*}
where we highlight the conditional structure of $Y$, given $\bX=\bx$.
Thus, in addition to the quantile regression of the previous sections, we aim at estimating 
regression functions $\bx \mapsto E_\tau^s(\bx)={\rm ES}^s_\tau(Y|\bx)$ for
$s \in \{-,+\}$.
Thanks to Theorem \ref{theo joint elicitability 2}, we know that we can jointly estimate
the triplet $(E^-_\tau(\bx), ~Q_{\tau}(\bx),~E^+_\tau(\bx))$ using the strictly consistent scoring function \eqref{eq:score}.
In analogy to \eqref{additive multiple quantile regression}, we choose  positively supported link functions $h$ and  $h_+$.
This motivates {\it deep composite model regression}
\begin{eqnarray}\nonumber
\bx & \mapsto& \left(E^-_\tau(\bx), ~Q_{\tau}(\bx),~E^+_\tau(\bx)\right)^\top
\\&&=\label{deep composite model regression}
\bigg( h^{-1}\langle \bbeta_{1}, \bz^{(d:1)}(\bx) \rangle, ~
h^{-1}\langle \bbeta_{1}, \bz^{(d:1)}(\bx) \rangle+h_+^{-1}\langle \bbeta_{2}, \bz^{(d:1)}(\bx) \rangle, 
\\&&\hspace{2cm}  h^{-1}\langle \bbeta_{1}, \bz^{(d:1)}(\bx) \rangle+
h_+^{-1}\langle \bbeta_{2} \bz^{(d:1)}(\bx) \rangle+h_+^{-1}\langle \bbeta_{3}, \bz^{(d:1)}(\bx) \rangle
 \bigg)^\top ~\in ~\R_+^3,\nonumber
\end{eqnarray}
for network parameter $\bvartheta=(\bw^{(1)}_1, \ldots, \bw^{(d)}_{r_d}, \bbeta_{1}, \bbeta_{2}, \bbeta_{3})^\top$.
Again, we choose a common deep FN network $\bz^{(d:1)}$ that is shared by the $\tau$-quantile
and the lower and upper ES.
Alternatively to \eqref{deep composite model regression}, we could also use a multiplicative approach similar to 
\eqref{multiplicative multiple quantile regression}.
We obtain the following M-estimator
\begin{eqnarray}\label{MLE NN ES}
&&\hspace{-2cm}
\widehat{\bvartheta}_\tau ~= ~\underset{\bvartheta}{\arg\min}~
\frac{1}{n} \sum_{i=1}^n 
 L\left(Y_i;
E^-_\tau(\bX_i), ~Q_{\tau}(\bX_i),~E^+_\tau(\bX_i)
\right),
\end{eqnarray}
where $L$ is a strictly consistent score given in \eqref{eq:score}.

\medskip

As already discussed at the end of Subsection \ref{subsec:choice}, in a regression context, the choice of the strictly consistent scoring function influences the (asymptotic) variance of the M-estimator. Therefore, we would like to provide some guidance on how to choose a score of the form \eqref{eq:score} in a data driven manner.
For simplicity, we will focus on scores of the form \eqref{eq:score additive decomp} and \eqref{eq:score revelation}, mainly discussing the choice of $\phi$, $\phi_-$ and $\phi_+$.
Moreover, motivated by modeling claim sizes, we assume that $Y>0$, a.s., such that $\Y = (0,\infty)$.

To this end, first recall a classical efficiency result in the context of mean regression; see Newey--McFadden \cite{NeweyMcFadden1994}. If $\mu(\bx) = \E[Y|\bX=\bx]$ is the estimable regression function and $\sigma(\bx)^2 = \mathbb V(Y|\bX=\bx)$ is the conditional variance, the most efficient Bregman score \eqref{Bregman divergence definition} should satisfy 
\begin{equation}
\label{eq:efficient Bregman}
\phi''\big(\mu(\bx)\big) = \frac{c}{\sigma(\bx)^2},
\end{equation}
for some $c>0$ and for all $\bx\in {\cal X}$. Since the third line of \eqref{eq:score revelation} is basically a Bregman score for the mean, we use condition \eqref{eq:efficient Bregman} to come up with a choice for $\phi$ in \eqref{eq:score revelation}.
For $\phi_+$ in \eqref{eq:score revelation} and \eqref{eq:score additive decomp}, we suggest to exploit a similar relation for the \emph{truncated} variance (recalling that $\ES_\tau^+$ is also a truncated mean). In particular, Theorem 4.3 in Dimitriadis et al.~\cite{DimitriadisFisslerZiegel2020} suggests 
\begin{equation}
\label{eq:efficient Bregman2}
\phi_+''\big(E_\tau^+(\bx)\big) = \frac{c_+}{\sigma_\tau^+(\bx)^2},
\end{equation}
for some $c_+>0$ and for all $\bx\in {\cal X}$, where $\sigma_\tau^+(\bx)^2=\mathbb V(Y| Y> F^{-1}_{Y|\bx}(\tau), \ \bX=\bx)$. Similarly,
\begin{equation}
\label{eq:efficient Bregman3}
\phi_-''\big(E_\tau^-(\bx)\big) = \frac{c_-}{\sigma_\tau^-(\bx)^2},
\end{equation}
for some $c_->0$ and for all $\bx\in {\cal X}$, where $\sigma_\tau^-(\bx)^2=\mathbb V(Y| Y\le F^{-1}_{Y|\bx}(\tau), \ \bX=\bx)$.

\medskip

To render this approach feasible, we suggest the following. 
Fit  a pre-estimate for the mean $\widehat \mu(\bx)$, using a strictly consistent score for mean estimation,
and an FN network for $\bx \mapsto \widehat \mu(\bx)$.
This allows us to study the squared Pearson's residuals, resulting in a non-parametric regression problem,
for $1\le i \le n$,
\begin{equation}
\label{eq:variance regression1}
\big(Y_ i - \widehat \mu(\bX_i)\big)^2 = \frac{c}{\phi''\big(\widehat \mu(\bX_i)\big)} + u_i, 
\end{equation}
where the error terms $u_i$ should be centered $\E[u_i|\bX_i]=0$. The goal is to solve 
\eqref{eq:variance regression1} for $c>0$ and $\phi''$. 
We suggest to simplify this problem and to turn it into a parametric problem by considering the following one-dimensional parametric family for $\phi$:
\begin{equation}
\label{eq:phi family}
\phi_b(y) = \begin{cases}
\frac{2}{b(b-1)}y^b, & \text{for }b\neq 0 \text{ and } b\neq 1,\\
-2\log(y), & \text{for }b=0,\\
2y\log(y)-2y, & \text{for }b=1,
\end{cases}
\end{equation}
where $y>0$. This provides us with Bregman divergences, see \eqref{Bregman divergence definition},
\begin{equation*}
L_{\phi_b}(y;a)=\phi_b(y)-\phi_b(a)-\phi_b'(a)(y-a)
= \begin{cases}
2\left[\frac{y^b}{b(b-1)}-y\frac{a^{b-1}}{b-1} +\frac{a^{b}}{b}\right], & \text{for }b\neq 0 \text{ and } b\neq 1,\\
2\left[\log(a/y)+(y-a)/a\right], & \text{for }b=0,\\
2\left[y\log(y/a)+a-y\right], & \text{for }b=1.
\end{cases}
\end{equation*}
For $b\not\in(1,2)$, these are exactly the deviance losses within Tweedie's family \cite{Tweedie} for power variance parameters $p=2-b$,
see Example 4.11 in W\"uthrich--Merz \cite{WM2021}. The case $b=2$ corresponds to the Gaussian
distribution, $b=1$ to the Poisson distribution,
$b=0$ to the gamma distribution and $b=-1$ to the inverse Gaussian distribution;
for $b \in (1,2)$ there are no Tweedie's distributions, see Theorem 2 in J{\o}rgensen \cite{Jorgensen2}. 
Thus, analyzing relation
\eqref{eq:variance regression1} will motivate the specific choice of $b$ and of $\phi_b$, respectively,
such that 
\begin{equation}\label{eq:variance regression 00}
\big(Y_ i - \widehat \mu(\bX_i)\big)^2 ~\approx~ \frac{c}{\phi_b''\big(\widehat \mu(\bX_i)\big)}
=\frac{c}{2}~\widehat \mu(\bX_i)^{2-b},
\end{equation}
and it allows us to select $c>0$.

For $\phi_+$ and $\phi_-$, we use a similar, though slightly more complicated, approach.
First, we come up with a pre-estimate for the conditional quantile function $\widehat Q_\tau(\bx)$, along the lines of Subsection \ref{subsec:Deep quantile regression}.
 Using this pre-estimate $\widehat Q_\tau(\bx)$, we split our sample into two distinct sets, introducing the index sets $\mathcal I_- = \{i\in \{1,\ldots, n\}| Y_i\le \widehat Q_\tau(\bX_i)\}$ and $\mathcal I_+ = \{i\in \{1,\ldots, n\}| Y_i > \widehat Q_\tau(\bX_i)\}$.
Fit conditional mean models $\widehat E_\tau^-(\bx)$ and $\widehat E_\tau^+(\bx)$ to the two data sets $\mathcal I_-$ and $\mathcal I_+$. 
Then, we can analyze
\begin{align}
\label{eq:variance regression2}
 \big(Y_ i - \widehat E_\tau^+(\bX_i)\big)^2 
&\approx \frac{c_+}{\phi''_+\big(\widehat E_\tau^+(\bX_i)\big)},\qquad i\in\mathcal I_+,\\
\label{eq:variance regression3}
\big(Y_ i - \widehat E_\tau^-(\bX_i)\big)^2 
&\approx \frac{c_-}{\phi''_-\big(\widehat E_\tau^-(\bX_i)\big)},\qquad i\in\mathcal I_-.
\end{align}
Again, one needs to estimate the parameters $c_-$ and $c_+$ as well as the functions $\phi''_-$ and $\phi''_+$,
and, for simplicity, we again suggest to use a member of the parametric family \eqref{eq:phi family},
so that we only need to select $b_-$, $c_-$, and $b_+$, $c_+$, respectively. 
Then, one obtains $\phi_-= \phi_{b_-}$ and $\phi_{+} = \phi_{b_+}$.
However, one should invoke the restrictions that $\phi'_{+}<0$ and $\phi'_{-}>0$. Therefore, we have the restriction $b_->1$ for $\phi_{b_-}$ and $b_+<1$ for  $\phi_{b_+}$.

\bigskip

We dispense with a discussion of the optimal choice of $g$. Equally so, we ignore conditions of the type of equation (4.19) in
Dimitriadis et al.~\cite{DimitriadisFisslerZiegel2020}, stipulating that for $G_{e^-, e^+}$ in \eqref{eq:increasing} it holds that 
$G'_{e^-, e^+}\big(Q_\tau(\bx)\big) = c_\tau f_{Y|\bx}\big(Q_\tau(\bx)\big)$
for some $c_\tau >0$ and for all $\bx\in\mathcal X$. Here, $f_{Y|\bx}$ is the conditional density of $Y$ given $\bX=\bx$.
Unless we use strong parametric assumptions on $f_{Y|\bx}$ to come up with a reasonable pre-estimate, we would need to resort to kernel density estimation methods here, which amounts to a considerable computational complexity.
We think that in the given setting, the precise estimation of the splicing point corresponding to the estimation of the conditional quantile is less important than the precise estimation of the lower and upper conditional ES, yielding a good estimate of the overall expected claim size.
Hence, we suggest to either set $g$ to be constant 0 or to use a multiple of the classical pinball loss, arising from $g(y) = c_\tau y$, $c_\tau>0$.

\medskip

We conclude that the overall claim size (pure risk premium) can be calculated by
\begin{equation}
\bx ~\mapsto ~
\E[Y|\bx]
= \tau {\rm ES}^-_\tau(Y|\bx) + (1-\tau) {\rm ES}^+_\tau(Y|\bx).
\end{equation}
The beauty of this approach now is that we can fit different models above and below the splicing point
$F_{Y|\bx}^{-1}(\tau)$, accounting for different properties in the main body and the tail of the data.
Thus, we can have different distributions above and below the splicing point, reflected in using different
scoring functions above and below $F_{Y|\bx}^{-1}(\tau)$  (implied by $\phi_-$ and $\phi_+$), resulting
in different regression functions potentially using covariates $\bx \in {\cal X}$ in different ways, i.e., we can
have different regression functions in the tail and the main body of the data.


\section{Real claim size example}
\label{Claim size example}
\subsection{Description of data}
\begin{figure}[htb!]
\begin{center}
\begin{minipage}[t]{0.4\textwidth}
\begin{center}
\includegraphics[width=.9\textwidth]{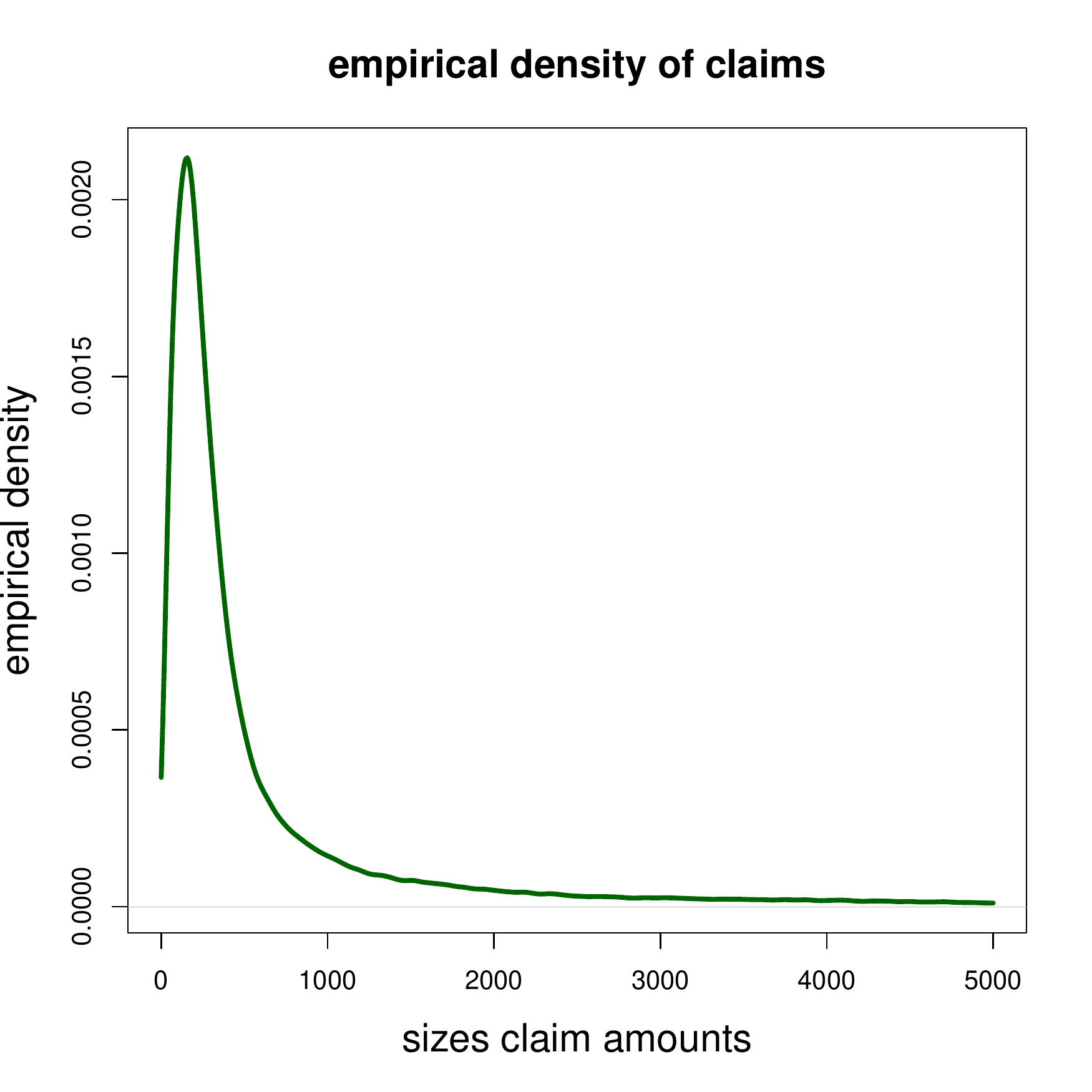}
\end{center}
\end{minipage}
\begin{minipage}[t]{0.4\textwidth}
\begin{center}
\includegraphics[width=.9\textwidth]{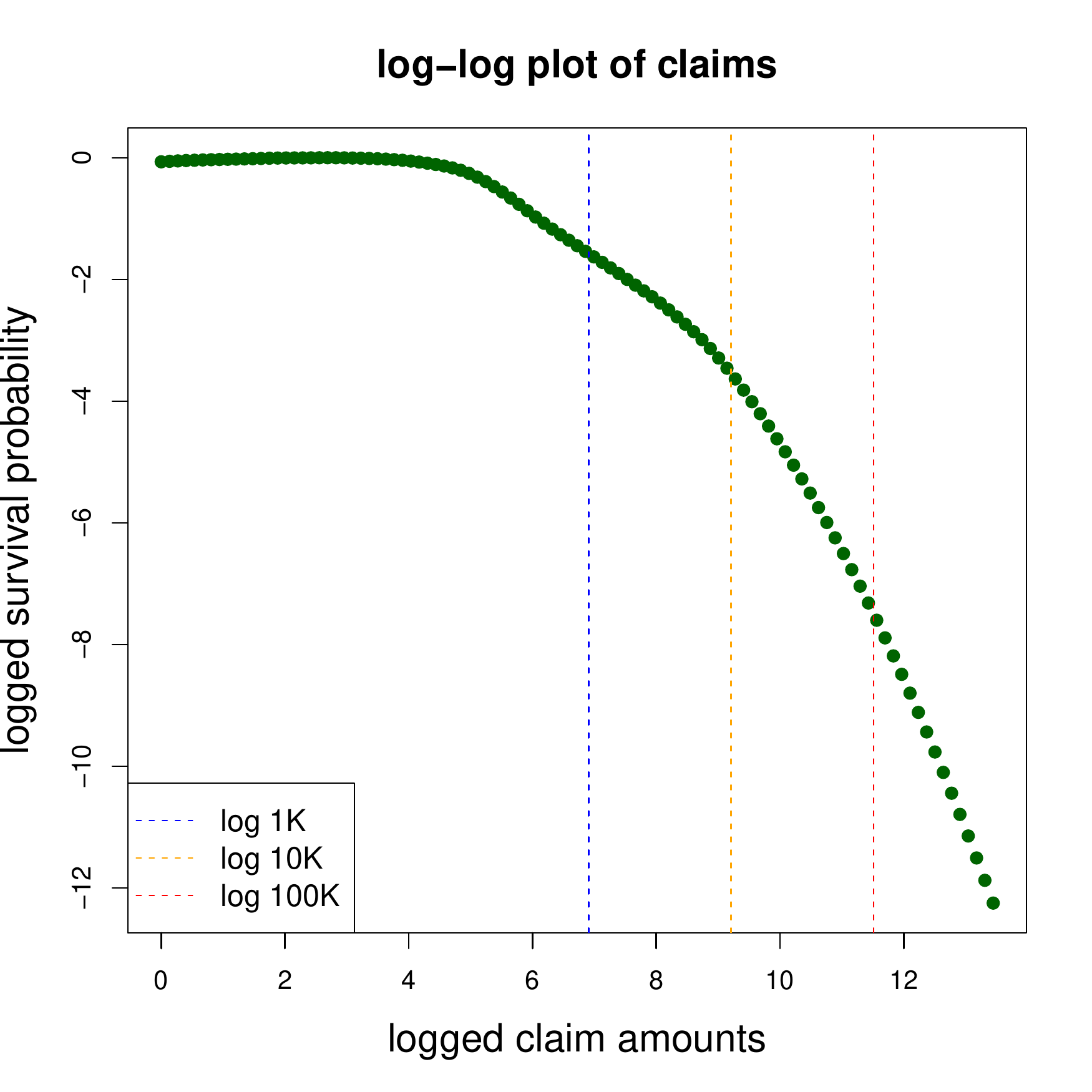}
\end{center}
\end{minipage}
\end{center}
\caption{(lhs) Empirical density (upper-truncated at 5,000), (rhs) log-log plot of observed Swiss
accident insurance claim amounts.}
\label{claim sizes UVG}
\end{figure}

We present a real data example with claim amounts describing medical expenses in 
compulsory Swiss accident insurance. In total we have
267,992 claims with positive claim amounts, $Y_i>0$, 
and these claim amounts range
from 1 to 691,066 CHF.  Figure \ref{claim sizes UVG} shows the empirical density and the log-log
plot of these  claim amounts. The empirical density is unimodal, and from the log-log plot
we conclude that the tail is moderately heavy-tailed, but it is not regularly varying (which
would correspond to an asymptotic straight line in the log-log plot).

These claim amounts are supported by 7 features. We have 3 categorical features
for the `labor sector' of the injured, the `injury type' and the `injured body part', 1 binary feature telling whether the injury is
a  `work or leisure' accident, and 3 continuous features corresponding to the `age' of the injured, the `reporting
delay' of the claim and the `accident quarter' (capturing seasonality in claims because leisure activities differ
in summer and winter times). A preliminary analysis shows that all these features have predictive power, i.e.,
they are explaining systematic effects in the claim amounts.

For our analysis, we partition the entire data into learning data ${\cal L}=(Y_i,\bx_i)_{1\le i \le n}$ that is used
for model fitting, and  test data ${\cal T}=(Y^\dagger_t, \bx^\dagger_t)_{1\le t \le T}$ 
which we (only) use for an out-of-sample
analysis. We do this partition stratified w.r.t.~the claim amounts and in a ratio of $9:1$. This results in learning data 
${\cal L}$ of size $n=241,193$ and in test data ${\cal T}$ of size $T=26,799$. We hold on to the same partition in all examples studied. For network fitting we further partition the learning data ${\cal L}=(Y_i,\bx_i)_{1\le i \le n}$
into training data ${\cal U}$ and validation data ${\cal V}$.
Thus, ${\cal U} \cup {\cal V}$ and ${\cal T}$ are disjoint (and assumed to be independent) so that we can
perform a proper out-of-sample forecast evaluation. For a detailed discussion of such a partition of
the data for SGD fitting we refer to Section 7.2.3 
and, in particular, to Figure 7.7 in W\"uthrich--Merz \cite{WM2021}.

\subsection{Example: deep multiple quantile regression}
\label{Example: deep multiple quantile regression}
We apply the framework of Section \ref{Deep multiple quantile regression} to perform a deep
multiple quantile regression. We choose 3 probability levels $0<\tau_1< \tau_2 < \tau_3<1$ that we 
simultaneously estimate; the specific choices considered are  
$(\tau_1,\tau_2,\tau_3) = (10\%, 50\%, 90\%)$.

We first discuss pre-processing of feature components before choosing the deep FN network architecture
$\bz^{(d:1)}$. For the 3 categorical variables we use embedding layers of dimension 2, i.e., they
are treated by a contextualized embedding. Embedding layers are explained in
Section 7.4.1 of W\"uthrich--Merz \cite{WM2021}. We use the {\sf R} library {\tt keras} for our 
implementation, and these embedding layers are encoded on lines 3-13 of Listing 
\ref{QuantAdd} in the supplementary material. The binary variable is encoded by 0-1 and the continuous variables
are pre-processed by the MinMaxScaler to ensure that they live on the same scale, see formula (7.30) in W\"uthrich--Merz \cite{WM2021} for the MinMaxScaler.
Based on this feature encoding we use an FN network of depth $d=3$ having input dimension
$r_0=3\cdot 2 + 1 + 3=10$ (for the categorical, binary and continuous feature components). This
 10-dimensional variable enters the network on line 15 of Listing \ref{QuantAdd}.
For the deep FN network  architecture $\bz^{(d:1)}$ we choose depth $d=3$ with $(r_1,r_2,r_3)=(20,15,10)$
hidden neurons in the hidden layers, and the hyperbolic tangent activation function $\Psi$. This is encoded
on lines 16-18 of Listing \ref{QuantAdd} and gives us learned representations $\bz_i =\bz^{(d:1)}(\bx_i)
\in \R^{r_d+1}$ of dimension $r_d+1=11$ of features $\bx_i$.
We remark that for insurance data of sample size of roughly 100,000 to 500,000 and with 10 to 20 feature
components we have had good experiences by an FN network architecture of this complexity, confirmed by the various examples in W\"uthrich--Merz \cite{WM2021}.
Therefore, we hold on to this choice.

Next we implement an additive structure \eqref{additive multiple quantile regression}
for deep multiple quantile regression
\begin{equation*}
\bx ~ \mapsto ~ \big(Q_{\tau_1}(\bx),~Q_{\tau_2}(\bx),~Q_{\tau_3}(\bx)\big)^\top ~\in~\R^3_+.
\end{equation*}

This requires that we (re-)use learned representations $\bz_i =\bz^{(d:1)}(\bx_i)$ 
in the last hidden layer three times and 
the sequence of quantiles should be monotonically increasing in $\tau_j$. We choose as link functions $h$ and $h_+$
the log-link which provides us with the exponential function for their inverses, and we then model
these quantiles recursively to preserve monotonicity. Lines 20-26 of Listing \ref{QuantAdd}
give the corresponding {\sf R} code, and line 28 outputs these ordered quantiles.
The multiplicative approach is rather similar, and the corresponding changes in the {\sf R} code are shown in 
Listing \ref{QuantMult} in the supplementary material.

This network architecture has $r=834$ network parameters that need to be fitted to the learning data ${\cal L}$.
We use pinball losses \eqref{pinball loss} for the probability
levels $(\tau_1,\tau_2,\tau_3) = (10\%, 50\%, 90\%)$; Listing \ref{QuantFitting} in the supplementary material shows the corresponding {\sf R} code.
These pinball losses then enter the compilation of the model on line 8
of Listing \ref{QuantFitting}. Moreover, we use the {\tt nadam} version of SGD which usually
has a good fitting performance. 
We fit the two architectures (additive and multiplicative) to our learning data ${\cal L}$, using 
80\% of the learning data ${\cal L}$ as training data ${\cal U}$ and 20\% as validation data ${\cal V}$
to explore the early stopping rule to prevent from over-fitting. 
To reduce the randomness of SGD fitting we repeat this procedure for 20
different starting points of the algorithm, and we calculate the average predictor over these 20 runs.
The results are given in Table \ref{results quantile regression}.

\begin{table}[htb!]
\centering
{\small
\begin{center}
\begin{tabular}{|l|ccc|}
\hline
&\multicolumn{3}{|c|}{out-of-sample pinball losses}\\
probability levels $\tau_j$ & $ 10\%$& $ 50\%$& $ 90\%$\\
\hline
additive approach& 141.69&  622.11&  717.33\\
multiplicative approach& 141.60  &622.72  &716.46\\
\hline
\end{tabular}

\end{center}}
\caption{Out-of-sample pinball losses $L_{\tau_j}$, $\tau_j \in \{10\%, 50\%, 90\%\}$, on the test data ${\cal T}$ of the deep 
multiple quantile regression using the additive and the multiplicative approaches.}
\label{results quantile regression}
\end{table}

Table \ref{results quantile regression} shows the average out-of-sample pinball losses 
$L_{\tau_j}$ on the test data ${\cal T}$ of the two
approaches, for $\tau_j \in \{10\%, 50\%, 90\%\}$. 
We note that the figures of the two approaches are very similar, and we cannot
give a clear preference to one of the two approaches. We further explore these results.

\begin{table}[htb!]
\centering
{\small
\begin{center}
\begin{tabular}{|l|ccc|}
\hline
&\multicolumn{3}{|c|}{out-of-sample coverage ratios}\\
probability levels $\tau_j$ & $ 10\%$& $ 50\%$& $ 90\%$\\
\hline
additive approach& 10.35\%& 50.56\%&  90.22\%\\
multiplicative approach& 10.39\%& 50.74\%&  90.25\%\\
\hline
\end{tabular}

\end{center}}
\caption{Out-of-sample empirical coverage ratios $\widehat{\tau}_j$ below the estimated deep quantile estimates
$Q_{\tau_j}(\bx^\dagger_t)$ for $\tau_j \in \{10\%, 50\%, 90\%\}$, see \eqref{quantile ratios}.}
\label{results quantile regression 2}
\end{table}

Table \ref{results quantile regression 2} shows the out-of-sample empirical coverage ratios on test data ${\cal T}$
of the event that the observation $Y_t^\dagger$ is smaller or equal to the estimated quantile, that is,
we evaluate
\begin{equation}\label{quantile ratios}
\widehat{\tau}_j~=~
\frac{1}{T} \sum_{t=1}^T \mathds{1}_{\left\{Y_t^\dagger \le Q_{\tau_j}(\bx^\dagger_t)\right\}},
\end{equation}
where $Q_{\tau_j}(\bx^\dagger_t)$ is the estimated quantile for level $\tau_j \in  \{10\%, 50\%, 90\%\}$ (using either the
additive or  the multiplicative approach) and evaluated in the features $\bx^\dagger_t$ of the out-of-sample
observations $Y_t^\dagger$, $1\le t \le T$. Table \ref{results quantile regression 2} verifies that the fitted deep 
multiple quantile
regression finds the quantiles (on portfolio level) very well because $\widehat{\tau}_j \approx \tau_j$;
we emphasize that these are out-of-sample figures.

\begin{figure}[htb!]
\begin{center}
\begin{minipage}[t]{0.6\textwidth}
\begin{center}
\includegraphics[width=.9\textwidth]{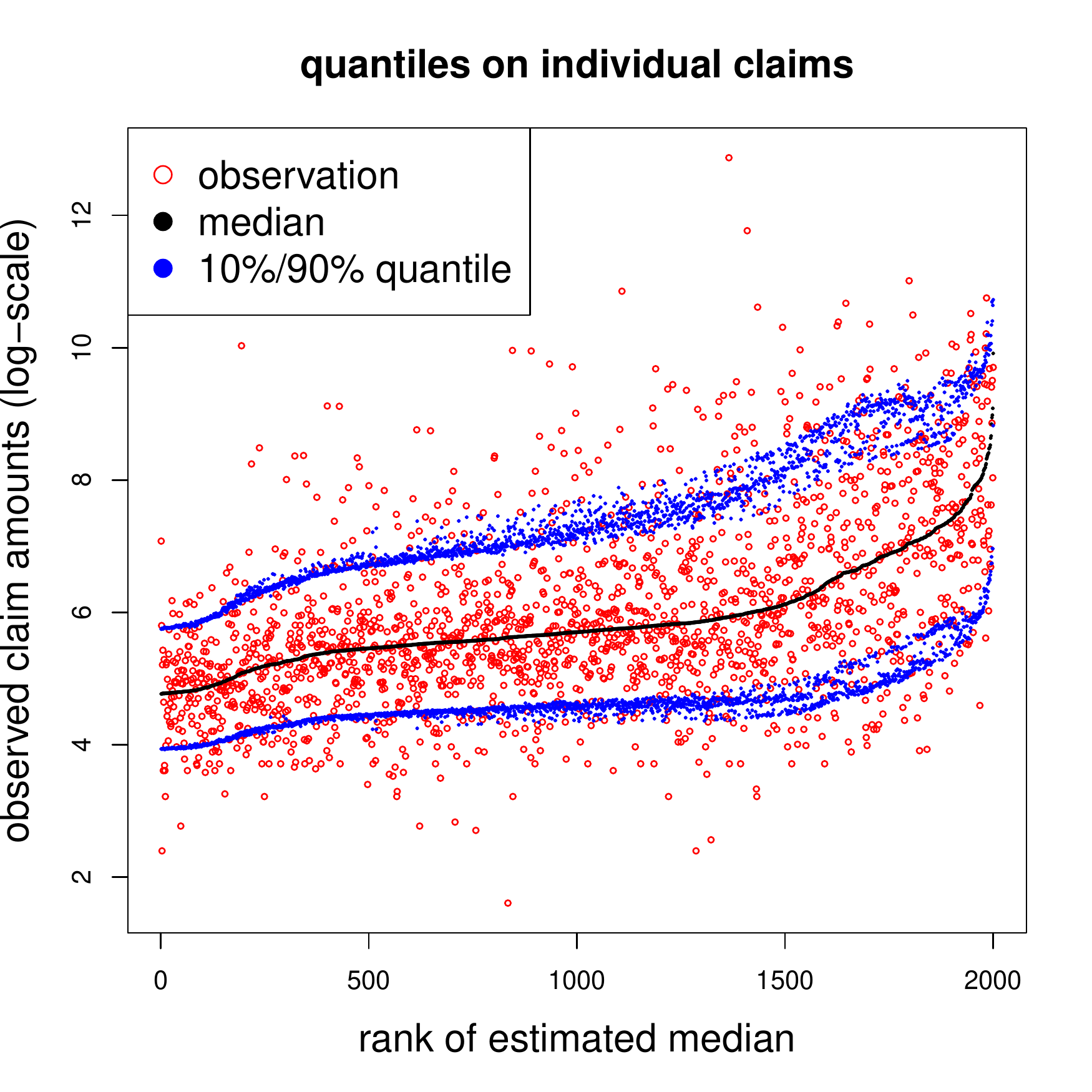}
\end{center}
\end{minipage}
\end{center}
\caption{Estimated quantiles $Q_{\tau_j}(\bx^\dagger_t)$ of 2,000 randomly 
selected individual features $\bx^\dagger_t$
on probability levels $\tau_j \in  \{10\%, 50\%,  90\%\}$ (blue, black, blue), and
the red dots show the corresponding out-of-sample observations (realizations) $Y^\dagger_t$; the $x$-axis orders the
claims w.r.t.~the estimated median  $Q_{50\%}(\bx_t^\dagger)$ (in black).}
\label{graph quantiles individual}
\end{figure}

Figure \ref{graph quantiles individual} shows the estimated quantiles
$Q_{\tau_j}(\bx^\dagger_t)$ for individual features $\bx^\dagger_t$
at probability levels $\tau_j \in \{10\%, 50\%,  90\%\}$ (blue, black, blue).
The individual observations are ordered w.r.t.~the estimated median in black. We observe that this ordering
does not imply monotonicity for the other quantiles, especially for bigger claim potentials. This indicates heteroskedasticity in our data. The red
dots show the corresponding (out-of-sample) observed claim amounts $Y^\dagger_t$.

We have analyzed these quantiles also in different granularity, for instance, we have analyzed the empirical
coverage ratios on feature levels. Also there the figures look good, but we refrain from giving more plots
because we would like to focus on the deep composite model regression proposed
in Section \ref{Deep multiple quantile regression}.

\subsection{Preliminary considerations for deep composite regression}
\label{Benchmark: deep double gamma regression model}
To fit a deep composite regression model we need a first preliminary step to explore relationship
\eqref{eq:variance regression 00}. This motivates the choices of $\phi$, $\phi_+$ and $\phi_-$ in 
\eqref{eq:Phi1}-\eqref{eq:Phi3}; for $g$ we use the identity function $g(y)=y$, giving us the pinball loss.
For this preliminary step, we choose exactly the same network architecture 
\eqref{quantile NNs} as for deep quantile regression,
the only change is that we replace the pinball loss by a strictly consistent scoring function
for the mean functional, see Theorem \ref{theorem Thomson}. As Bregman divergence we choose
the gamma deviance loss, which corresponds to choice $b=0$ in \eqref{eq:phi family}. We remark that
the gamma model is the most popular model for insurance claim size modeling, and it often provides
a first reasonable choice for a regression model; this is also the case for this preliminary step.

We fit three networks of depth $d=3$ with $(r_1,r_2,r_3)=(20,15,10)$ neurons and exponential output activation $h^{-1}$
to (a) all learning data ${\cal L}$, (b) the learning data having observations $Y_i > 
Q_{90\%}(\bx_i)$, and (c) the learning data having observations $Y_i \le 
Q_{90\%}(\bx_i)$, where $Q_{90\%}(\bx_i)$ is the estimated $\tau=90\%$
quantile from the previous example of Section \ref{Example: deep multiple quantile regression}.
This allows us to analyze \eqref{eq:variance regression1} providing a choice for $\phi$, 
\eqref{eq:variance regression2} giving a choice for $\phi_+$, and
\eqref{eq:variance regression3} giving a choice for $\phi_-$ for a deep composite model regression
at probability level $\tau=90\%$.

\begin{figure}[htb!]
\begin{center}
\begin{minipage}[t]{0.32\textwidth}
\begin{center}
\includegraphics[width=.9\textwidth]{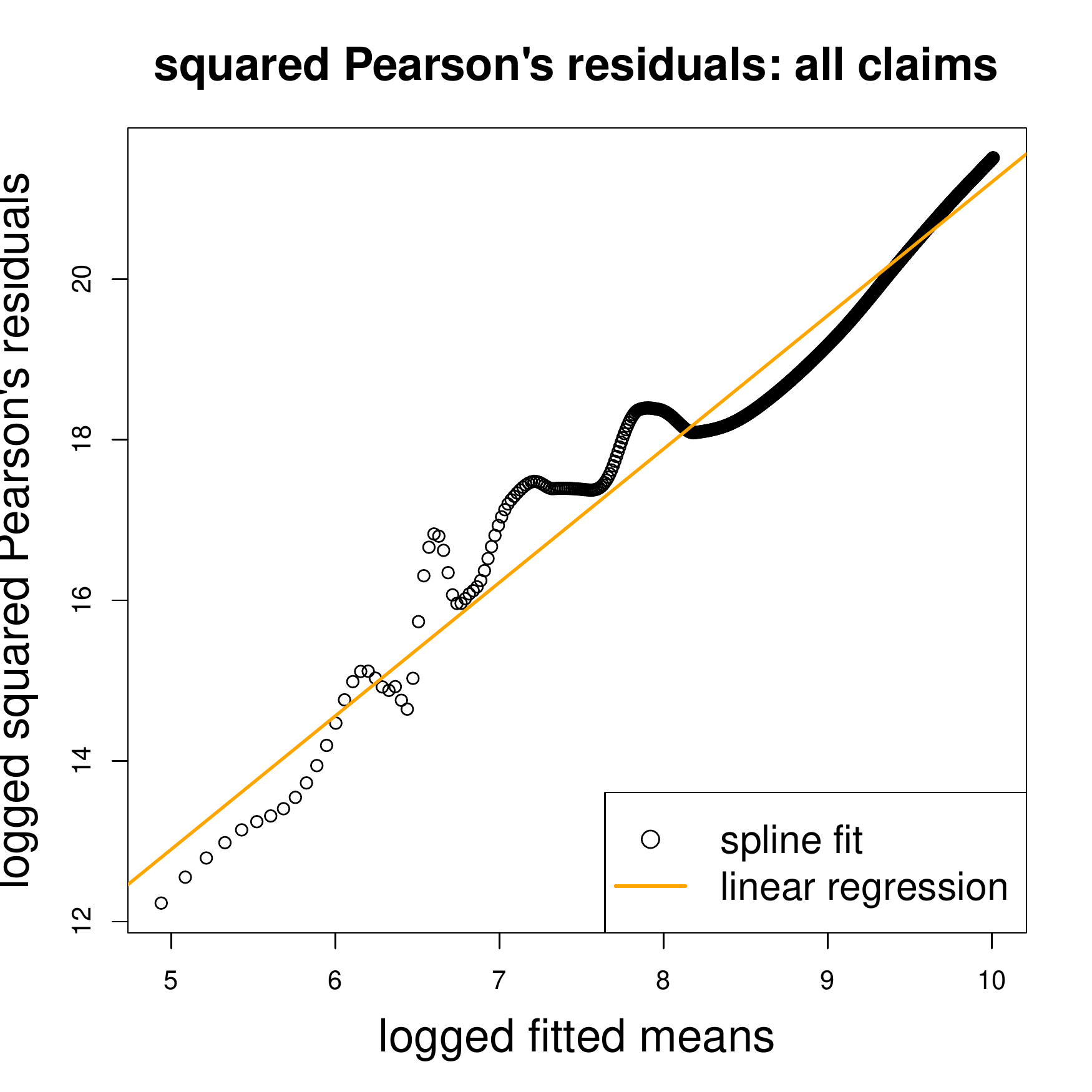}
\end{center}
\end{minipage}
\begin{minipage}[t]{0.32\textwidth}
\begin{center}
\includegraphics[width=.9\textwidth]{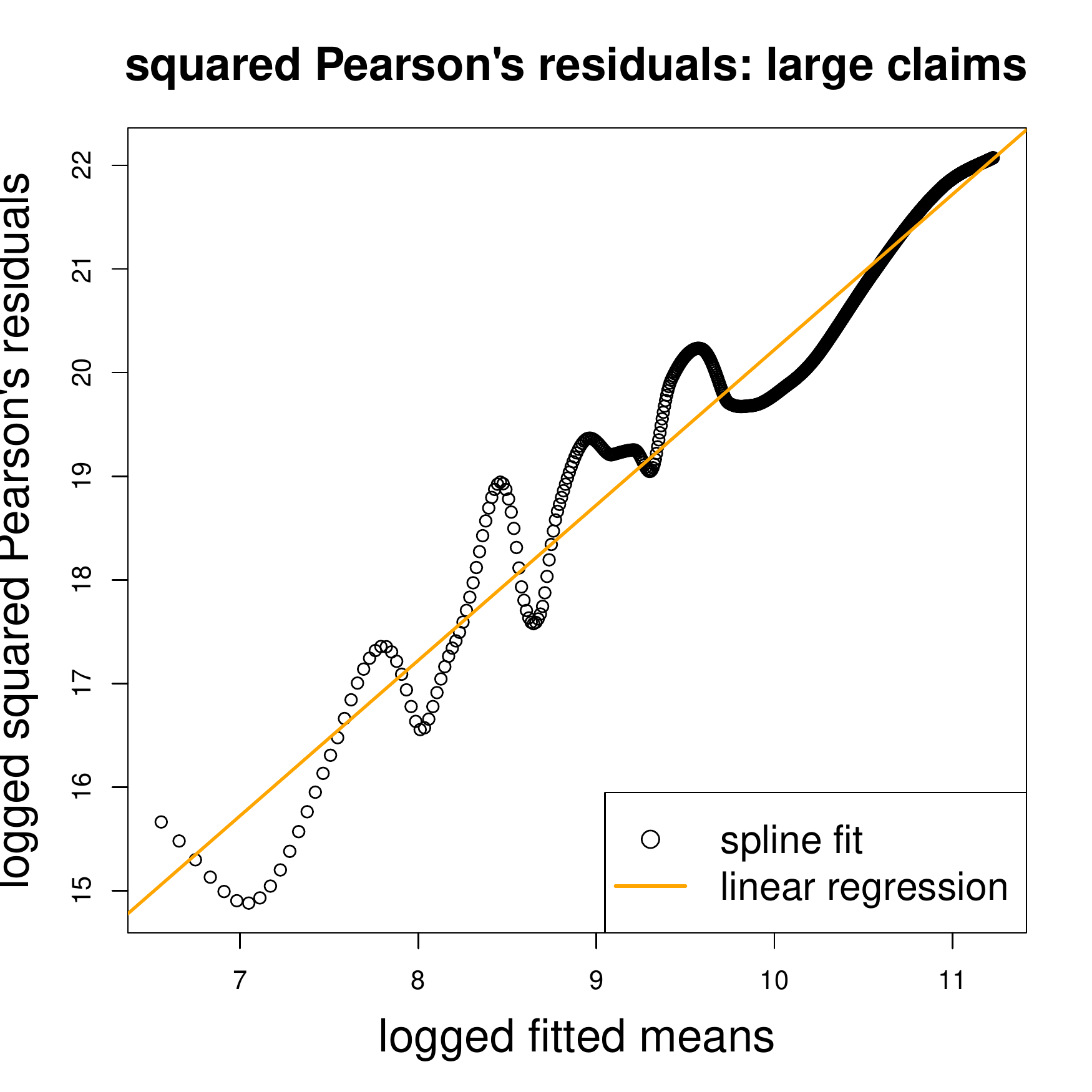}
\end{center}
\end{minipage}
\begin{minipage}[t]{0.32\textwidth}
\begin{center}
\includegraphics[width=.9\textwidth]{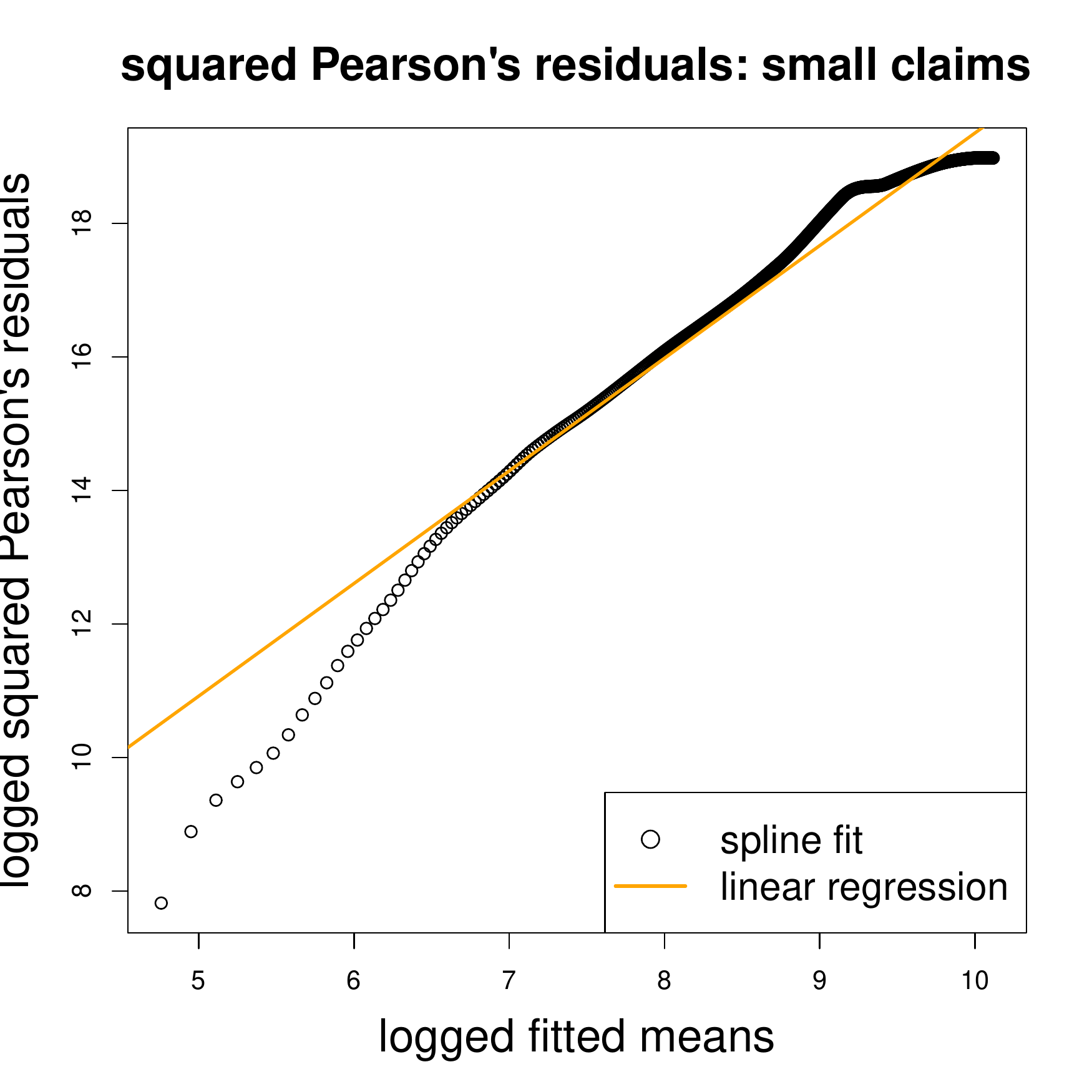}
\end{center}
\end{minipage}
\end{center}
\caption{Spline fit and linear regression to the estimated squared Pearson's residuals as a function of the
estimated means (both axis are on the log-scale): (lhs) all claims, (middle) large claims above the 90\%-quantile,
and (rhs) small claims below the 90\%-quantile.}
\label{determination of b}
\end{figure}

Figure \ref{determination of b} shows a spline fit and a linear regression to the
estimated squared Pearson's residuals $(Y_i - \widehat{\mu}(\bx_i))^2$, where $\widehat{\mu}(\cdot)$ is the 
estimated mean functional and where both axis are on the log-scale. 
The left-hand side shows all claims, the middle shows the situation
where we only fit the network to the claims above the estimated quantile $Q_{90\%}(\bx_i)$,
and the right-hand side only considers the claims below that quantile. 

\begin{table}[htb!]
\centering
{\small
\begin{center}
\begin{tabular}{|l|ccc|}
\hline
& all claims & large claims & small claims\\
&$\phi$ & $\phi_+$& $\phi_-$\\
\hline
intercept $\log(c/2)$& 4.592 & 5.229 & 2.483\\
slope $2-b$ & 1.662  &  1.499& 1.687\\
parameter $b$ & 0.338  & 0.401& 0.313\\
\hline
\end{tabular}
\end{center}}
\caption{Linear regression parameters for $c\phi_b(\cdot)/2$ from Figure \ref{determination of b}.}
\label{linear regression parameters}
\end{table}

Table \ref{linear regression parameters}
gives the linear regression estimates for $c$ and $b$ in \eqref{eq:variance regression 00} for the three cases.
We observe that in all three cases we receive $b<1$ which results in derivatives $\phi'_b(y)=y^{b-1}/(b-1)<0$. This
implies that we can only work under scoring function \eqref{eq:score revelation}, because $\phi_-$ requires
$b>1$, e.g., the square loss function with $b=2$ would work for $\phi_-$ but this will not provide
optimal convergence rates according to \eqref{eq:variance regression3}.

\subsection{Deep composite model regression}
We are now ready to fit the deep composite regression model for the claims $Y$ w.r.t.~the probability level $\tau=90\%$.
To this end, we implement the regression function \eqref{deep composite model regression} and as deep 
FN network we use the same architecture as in the additive deep multiple quantile regression approach.
This regression model is then fitted under the strictly consistent scoring function \eqref{eq:score revelation}
for the joint $\tau$-quantile and lower and upper ES (using the parameters of Table \ref{linear regression parameters}).

\begin{figure}[htb!]
\begin{center}
\begin{minipage}[t]{0.45\textwidth}
\begin{center}
\includegraphics[width=.9\textwidth]{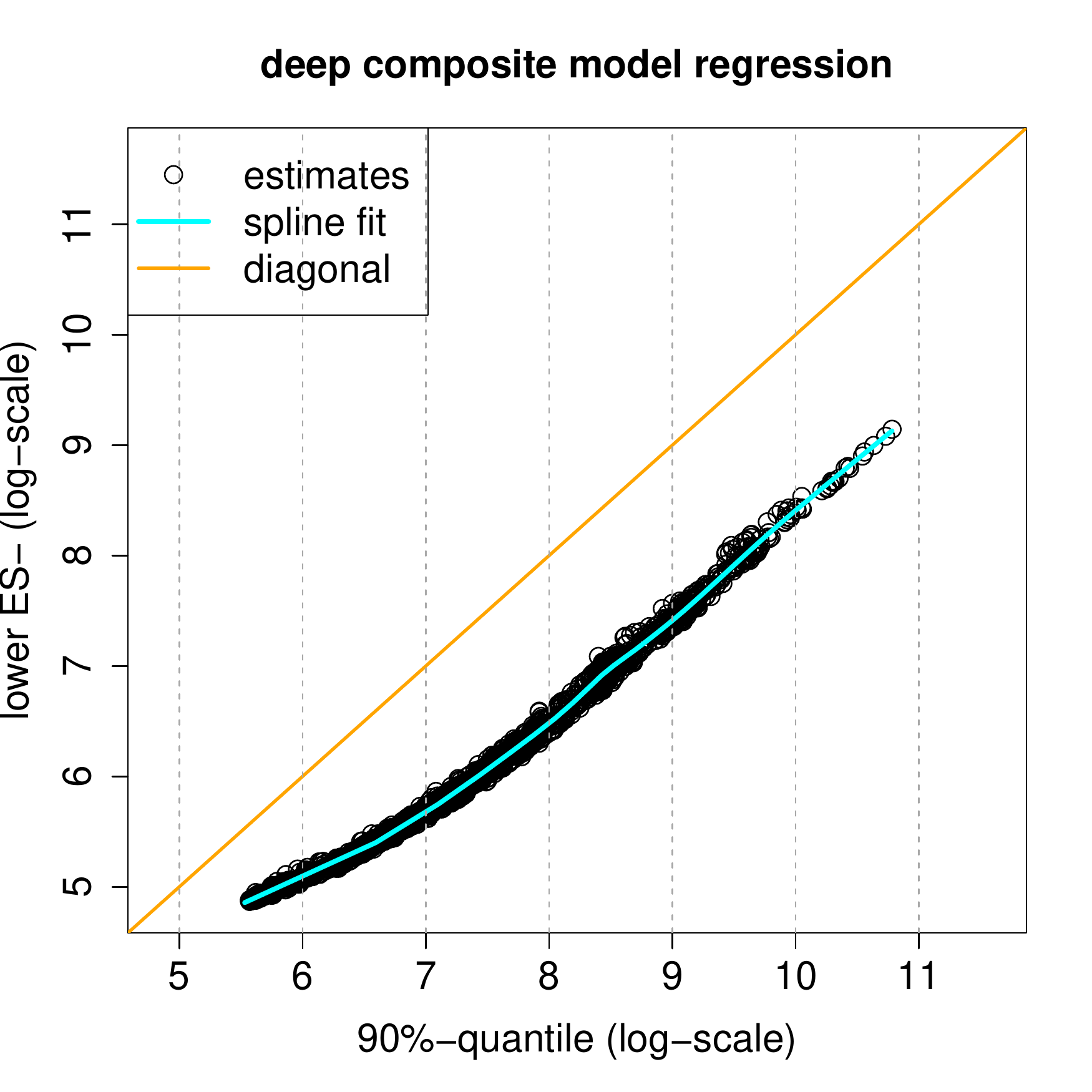}
\end{center}
\end{minipage}
\begin{minipage}[t]{0.45\textwidth}
\begin{center}
\includegraphics[width=.9\textwidth]{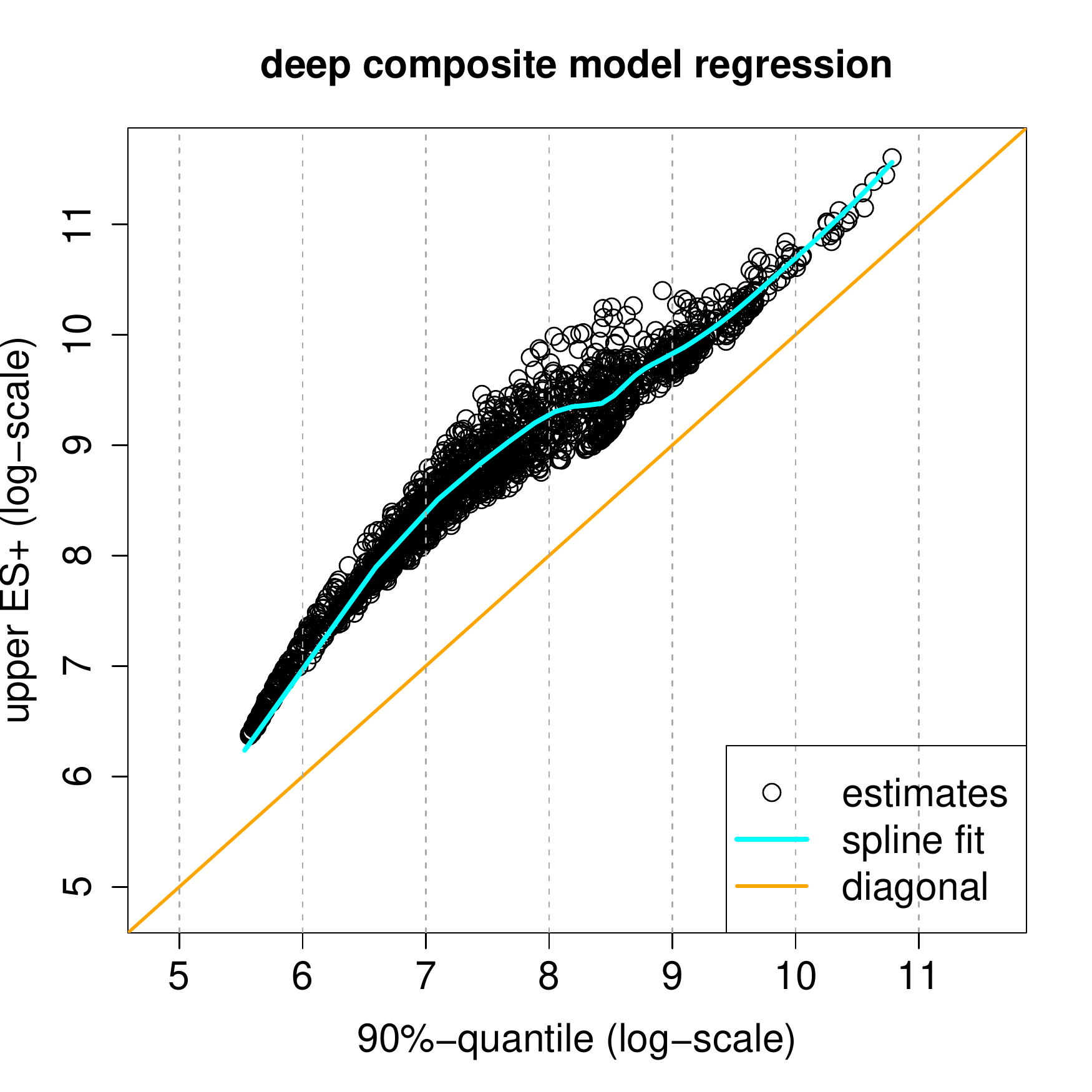}
\end{center}
\end{minipage}
\end{center}
\caption{(lhs) Estimated lower $E^-_{90\%}(\bx_t^\dagger)$ vs.~estimated quantile 
$Q_{90\%}(\bx_t^\dagger)$
and (rhs) estimated upper $E^+_{90\%}(\bx_t^\dagger)$
vs.~estimated quantile 
$Q_{90\%}(\bx_t^\dagger)$
at probability level $\tau=90\%$.}
\label{expected shortfall plot}
\end{figure}

We fit this network under scoring function \eqref{eq:score revelation}, which is encoded in 
Listing \ref{CompositeFitting}, using the {\tt nadam} version of SGD.
Figure \ref{expected shortfall plot} shows the out-of-sample estimated lower ES,
$E^-_{90\%}(\bx_t^\dagger)$, and upper ES, $E^+_{90\%}(\bx_t^\dagger)$, against the estimated quantiles, $Q_{90\%}(\bx_t^\dagger)$, of 2,000 randomly selected instances 
$\bx_t^\dagger$, and the cyan lines present spline fits to all out-of-sample instances.
Basically, the gaps between the cyan lines and the diagonal orange line describe the differences between
the ES and the quantile. This gap is of constant size between the lower ES and the quantile above 7 (on the
log-scale) which means that the lower ES is a fixed ratio of the quantile. The structure of the upper ES
relative to the quantile is more complicated as the gap is becoming smaller with bigger values
for the 90\% quantile.

Based on these estimates we can now determine the expected value of $Y$, given $\bx$, 
\begin{equation*}
\widehat{\mu}(\bx) = \tau \widehat{\rm ES}^-_{\tau}(Y|\bx)+
(1-\tau) \widehat{\rm ES}^+_{\tau}(Y|\bx).
\end{equation*}
We compare these estimated means to the ones obtained from the (plain vanilla) deep gamma model used
in the preparatory Section \ref{Benchmark: deep double gamma regression model}; note that the
minimization of the gamma deviance loss with $b=0$ in \eqref{eq:phi family} is equivalent to the maximization
of the log-likelihood function of the gamma distribution.

\begin{figure}[htb!]
\begin{center}
\begin{minipage}[t]{0.45\textwidth}
\begin{center}
\includegraphics[width=.9\textwidth]{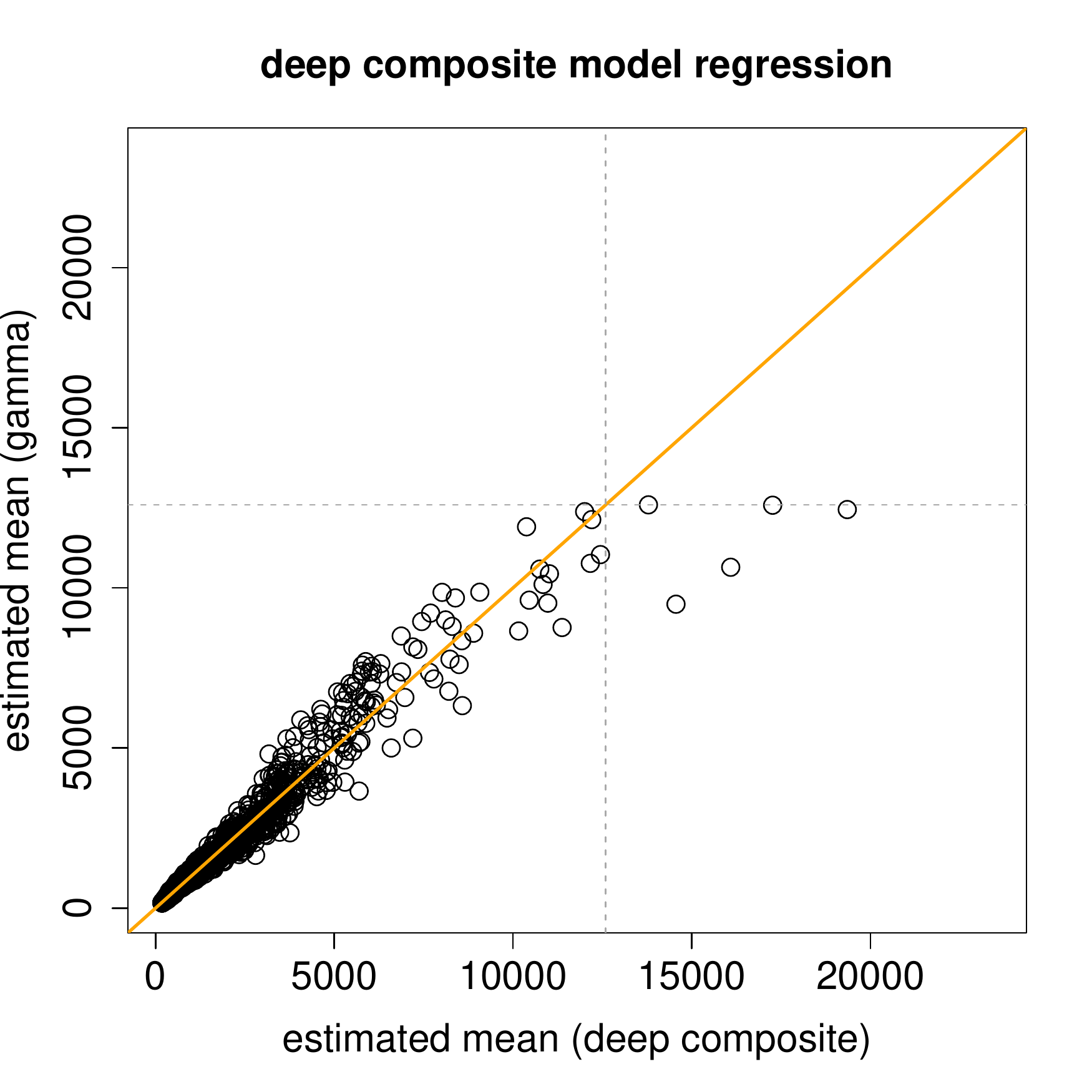}
\end{center}
\end{minipage}
\begin{minipage}[t]{0.45\textwidth}
\begin{center}
\includegraphics[width=.9\textwidth]{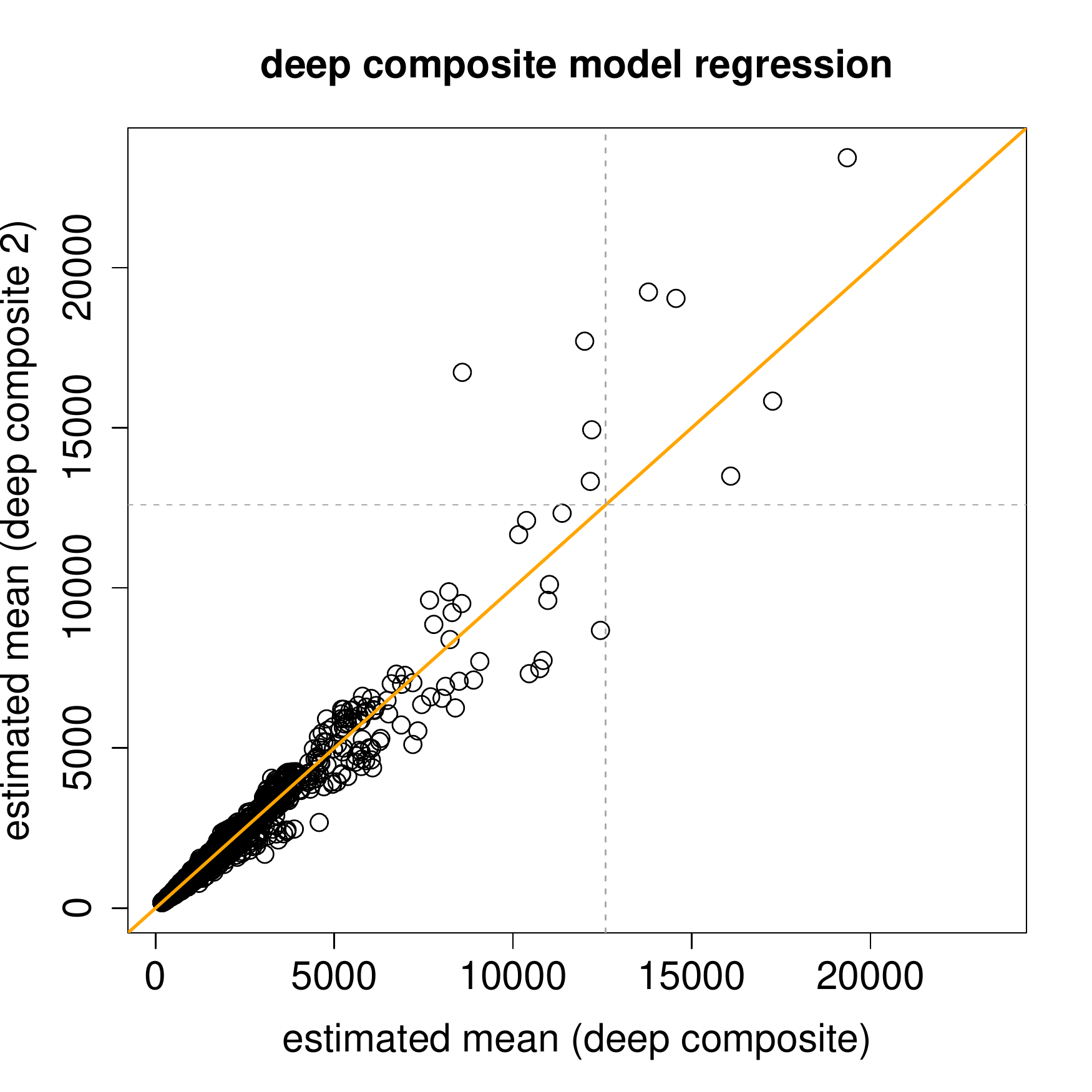}
\end{center}
\end{minipage}
\end{center}
\caption{Comparison of  estimated out-of-sample means in (lhs) the deep gamma
model and the deep composite model, and (rhs) the deep composite model 2 and
the deep composite  model.}
\label{estimated means}
\end{figure}

Figure \ref{estimated means} (lhs) compares the deep gamma model to the fitted deep composite
model. For small estimated means $\mu(\bx)$ the deep gamma and the deep composite model are rather similar,
however, for large estimated means the deep gamma model provides clearly smaller estimates, see gray dotted
lines in Figure \ref{estimated means} (lhs). It seems that the deep gamma model under-estimates claims with
large expected payments. To verify this, and to check the robustness of the results, we fit a second
deep composite model to the data. To this end, we choose a different strictly consistent scoring function,
namely, we choose \eqref{eq:score additive decomp} which separates small and large claims in an additive
way through $\phi_-$ and $\phi_+$. For the main body of the claims (below the 90\%-quantile) we choose
the square loss function for $\phi_-$ (which corresponds to choice $b=2$ in \eqref{eq:phi family}), and for the claims above the
probability level we choose $b=0$ for $\phi_+$ in \eqref{eq:phi family}, which corresponds to the gamma deviance loss. We call this second model
`deep composite model 2'.
 
Figure \ref{estimated means} (rhs) compares the two fitted deep composite models, using different score functions
$L$, which implicitly implies different distributional assumptions in an MLE context. We observe that under both
score functions we receive rather similar results, which verifies the robustness of the fittings. There are differences
because SGD fitting with early stopping typically finds different `good' solutions, and this inherent fluctuation
in SGD can
only be reduced by averaging (blending) over model calibrations.

\begin{table}[htb!]
\centering
{\small
\begin{center}
\begin{tabular}{|l|ccc|}
\hline
& coverage & lower ES & upper ES \\
&  ratio & identification & identification\\
& $\widehat{\tau}=90\%$ & $\widehat{v}_-$ & $\widehat{v}_+$ \\\hline
deep composite model & 90.13\% &1.3&-170.7\\
deep composite model 2 & 90.10\% &-16.5&-203.9\\
deep gamma model & 93.63\% &5,612.7&-7,962.4
\\\hline
\end{tabular}
\end{center}}
\caption{Out-of-sample empirical coverage ratios $\widehat \tau$ and
identification functions $\widehat v_-$ and $\widehat v_+$ of the three considered models.}
\label{comparison 4 a}
\end{table}

Analogously to checking the empirical out-of-sample coverage ratios in Table \ref{results quantile regression 2}, which validated the calibration of our deep quantile models, we now check calibration for the composite triplet consisting of the quantile, the lower ES and the upper ES. 
For the calibration of $Q_{90\%}$ we again use the empirical coverage defined in \eqref{quantile ratios}, which should be close to $90\%$.
Unfortunately, the lower and the upper ES not only fail to be elicitable, but also turn out not to be identifiable; see Supplement \ref{subsec:Identifiability results} for a brief discussion of identifiability. 
Hence, we cannot check the calibration of the models $E_{90\%}^-$ and $E_{90\%}^+$ standalone, but we can only evaluate the goodness-of-fit of these models \emph{jointly} with the corresponding quantile model $Q_{90\%}$. 
The corresponding joint strict identification functions are given as the first and third component in \eqref{eq:V}.
The empirical out-of-sample identification functions are
\begin{align}
\label{eq:calibration1}
\widehat v_- &= \frac{1}{T} \sum_{t=1}^T \left[E^-_{90\%}(x_t^\dagger) - \frac{Y_t^\dagger}{0.9}\mathds{1}_{\{Y_t^\dagger \le Q_{90\%}(\bx_t^\dagger)\}} + \frac{Q_{90\%}(\bx_t^\dagger)}{0.9} \Big(\mathds{1}_{\{Y_t^\dagger \le Q_{90\%}(\bx_t^\dagger)\}} - 0.9\Big)\right],\\
\label{eq:calibration2}
\widehat v_+ &= 
\frac{1}{T} \sum_{t=1}^T \left[E^+_{90\%}(x_t^\dagger) 
- \frac{Y_t^\dagger}{0.1}\mathds{1}_{\{Y_t^\dagger> Q_{90\%}(\bx_t^\dagger)\}} 
- \frac{Q_{90\%}(\bx_t^\dagger)}{0.1} \Big(0.1 - \mathds{1}_{\{Y_t^\dagger > Q_{90\%}(\bx_t^\dagger)\}} \Big)
\right].
\end{align}
Values of $\widehat v_-$ and $\widehat v_+$ close to 0 indicate well calibrated models for $(E_{90\%}^-, Q_{90\%})$ and $(E_{90\%}^+, Q_{90\%})$, respectively.

Table \ref{comparison 4 a} reports the empirical coverage ratios $\widehat \tau$, for $\tau=90\%$, along with the empirical identification functions $\widehat v_-$ and $\widehat v_+$ for the two deep composite models and for the deep gamma model;
having a gamma distributional assumption from the deep gamma approach 
of Section \ref{Benchmark: deep double gamma regression model} we can also calculate the composite
triplet under this gamma model assumption. In Section \ref{Benchmark: deep double gamma regression model} 
we have fitted the means $\widehat{\mu}(\bx^\dagger_t)$ under the gamma assumption. Moreover, using these means
we can estimate a dispersion parameter. If we use the deviance dispersion estimate we receive in this gamma model
a dispersion parameter of $1/\gamma=1.83$. This allows us to calculate the corresponding quantiles and ES in the gamma model.
The lower and upper ES in the gamma model are obtained by
\begin{eqnarray*}
\E\left[Y\left|Y\le\Gamma_{\gamma, \mu}^{-1}(\tau)\right.\right] &=& \mu \left(\frac{\Gamma_{\gamma+1, \mu}\left(\Gamma_{\gamma, \mu}^{-1}(\tau)\right)}
{\tau}\right),\\
\\
\E\left[Y\left|Y>\Gamma_{\gamma, \mu}^{-1}(\tau)\right.\right] &=& \mu \left(\frac{1-\Gamma_{\gamma+1, \mu}\left(\Gamma_{\gamma, \mu}^{-1}(\tau)\right)}
{1-\tau}\right),
\end{eqnarray*}
where $Y\sim \Gamma_{\gamma, \mu}$ denotes the gamma distribution with mean $\mu>0$ and shape parameter $\gamma>0$.

From Table \ref{comparison 4 a} we conclude that the deep composite regression models meet the
right coverage ratio of 90\% very well, whereas the deep gamma model results in a too high coverage ratio which indicates
that the gamma distributional assumption does not match the tail of the observed data. This carries over to the identification
functions of the lower and upper ES. 
The first deep composite model seems to be slightly better calibrated than the second one, which is indicated by values of $\widehat v_-$ and $\widehat v_+$ closer to zero for the first model.
The deep gamma model exhibits a poor calibration reflected in high absolute values of $\widehat v_-$ and $\widehat v_+$.

\begin{figure}[htb!]
\begin{center}
\begin{minipage}[t]{0.32\textwidth}
\begin{center}
\includegraphics[width=.9\textwidth]{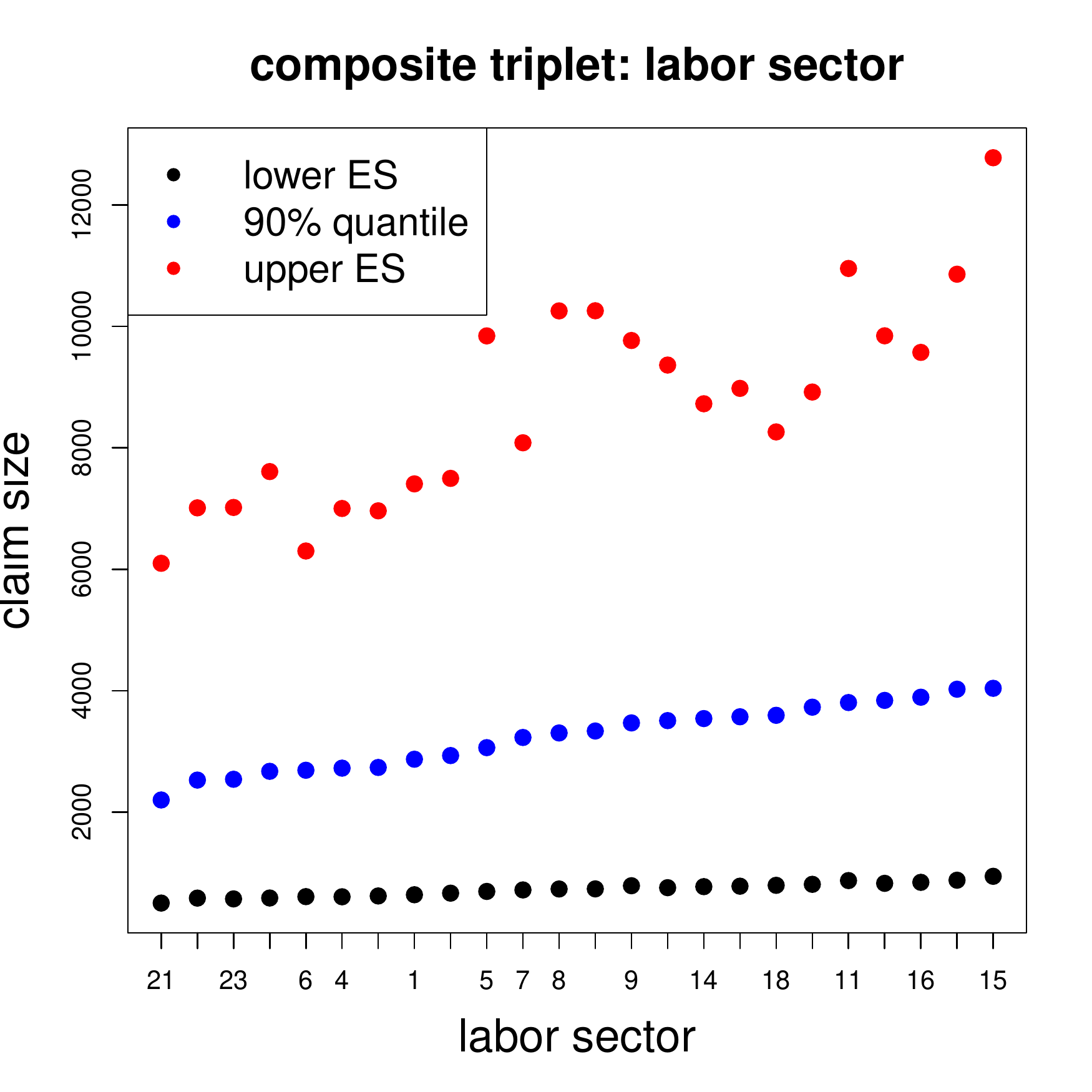}
\end{center}
\end{minipage}
\begin{minipage}[t]{0.32\textwidth}
\begin{center}
\includegraphics[width=.9\textwidth]{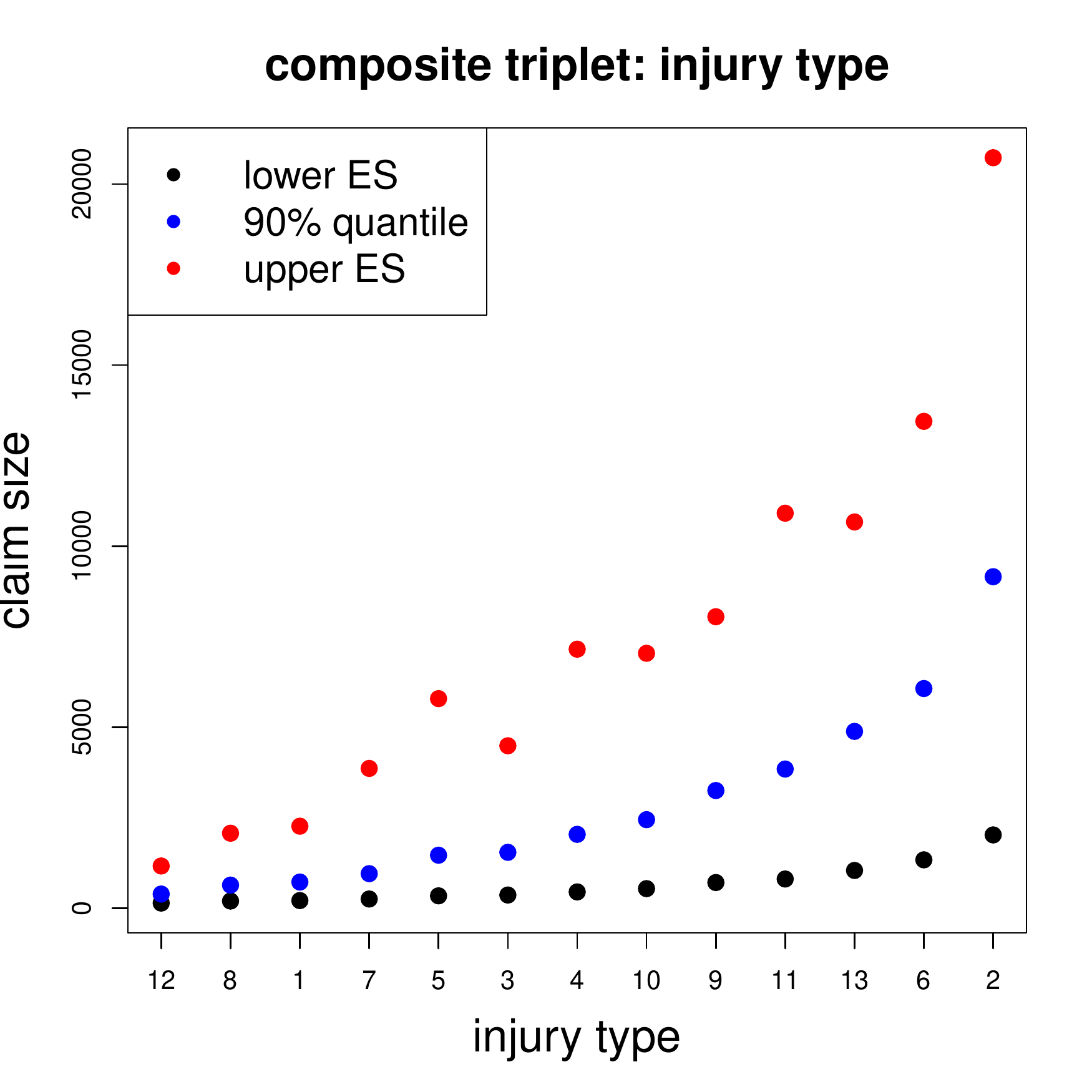}
\end{center}
\end{minipage}
\begin{minipage}[t]{0.32\textwidth}
\begin{center}
\includegraphics[width=.9\textwidth]{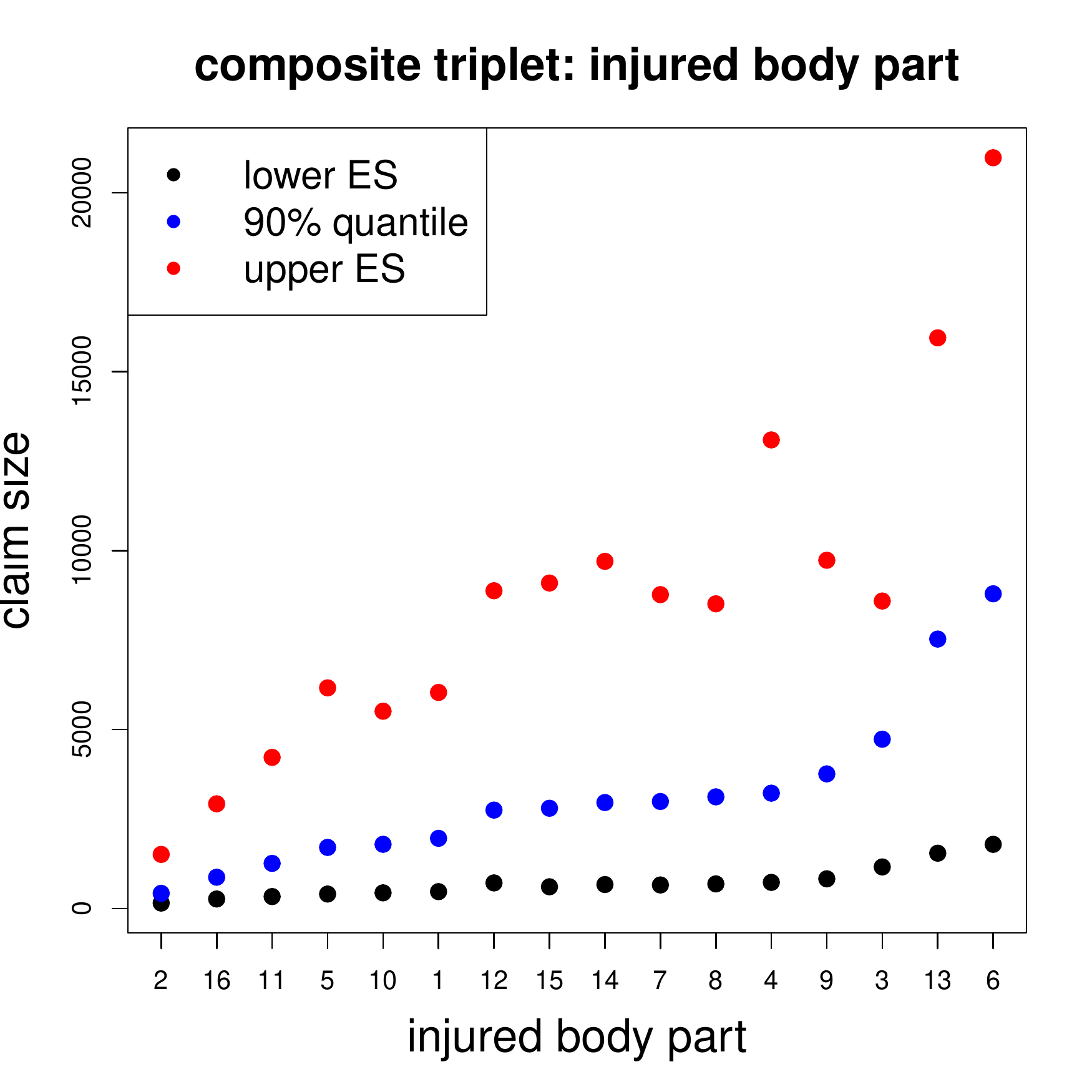}
\end{center}
\end{minipage}
\end{center}
\caption{Average marginal estimations for selected feature levels: (lhs) labor sector, (middle) injury type, (rhs)
injured body part; ordered on the $x$-axis w.r.t.~the average quantile estimates $Q_{90\%}$.}
\label{marginal predictions}
\end{figure}

Figure \ref{marginal predictions} gives marginal estimations of the 
composite triplet of lower ES, the 90\% quantile and the upper ES for the features `labor sector', `injury type' and `injured body part'; the $x$-axis is ordered w.r.t.~the average 90\% quantiles.
We observe monotonicity between the average lower ES and the $90\%$ quantiles (black and blue dots), but this monotonicity
gets lost w.r.t.~the upper ES (red dots). This indicates that we have different regression functions for the main body
and the tail of the data. That is, we have developed a very flexible (deep regression) model where features $\bx \in {\cal X}$ impact
predictions differently in the body and the tail of the distributions. This concludes the example.

\section{Summary and outlook}
\label{Conclusions}

We present deep quantile and deep composite model regression. 
For the composite model, we use a conditional quantile as splicing point, not an absolute threshold. This provides flexibility e.g.~in the presence of heteroskedasticity.
We utilize a network architecture which respects the natural ordering of quantiles at different probability levels as well as the natural ordering of the composite triplet consisting the the lower expected shortfall, the quantile and the upper expected shortfall at the same probability level.
While strictly consistent scoring functions for tuples of quantiles are available, e.g.~in the form of sums of pinball losses, and thus M-estimation can be performed for deep quantile models, we first derive and introduce the class of strictly consistent scoring functions for the composite triplet.
Addressing the specific choice of the strictly consistent score for the composite triplet, we discuss data-driven choices which potentially increase the efficiency in estimation.
The suitability of our methods is illustrated on a real data example with claim amounts describing medical expenses in compulsory Swiss accident insurance.

A relevant extension of our methods is a generalized composite model using two (or even more) splicing points. This allows
one to model the influence of features differently for small claim sizes, large claim sizes and the body of the data.
This extension seems particularly beneficial since there is empirical evidence that the (conditional) distribution of claim sizes is different in nature for small, claim sizes, the body, and for large claim sizes. Hence, this extension has the potential to increase accuracy of overall claim modeling, which is relevant in insurance pricing.
The positive results on the elicitability of the range value at risk (RVaR) -- or interquantile expectation -- together with the two corresponding quantiles in Fissler--Ziegel \cite{FisslerZiegel2021} suggest the existence of strictly consistent scoring functions for the ``extended composite quintuple'' consisting of two quantiles (say, the 10\% and 90\% quantiles) together with the lower ES, upper ES and the range value at risk.
However, the specific data-driven choice of the score motivated by efficiency considerations is far from being clear since the class of scoring functions for RVaR and two quantiles is less flexible than the one involving ES and a quantile. E.g.~it is shown in \cite{FisslerZiegel2021} that there are no positively homogeneous strictly consistent scores for the former case.
Therefore, we defer a detailed study of this extension to future research.

\section*{Declaration of competing interest}

There is no competing interest.

{\small 
\renewcommand{\baselinestretch}{.51}
}

\appendix
\newpage
\section{Supplementary material:\\
Characterisation results on scoring functions}
\label{Suppl:Characterisation}
\subsection{Identifiability results}
\label{subsec:Identifiability results}

Identification functions, also known as moment functions in econometrics, are closely related to scoring functions. Here, the correctly specified functional forecast annihilates the expected identification function rather than minimizing it.
Intuitively speaking and subject to smoothness conditions, identification functions can arise as gradients of scoring functions, which explains why their dimension commonly coincides with the dimension of the forecasts.

\begin{defi}[identification function]
Let $\A\subseteq \R^k$. A measurable function $V: \Y \times \A \to \R^k$ is a strict $\F$-identification function for a 
given functional
$A: {\cal F} \to {\cal P}(\A)$
if $\E_F \left[ |V_m(Y;a)| \right]<\infty$ for all $m\in\{1, \ldots, k\}$, for all $Y \sim F \in {\cal F}$ and for all $a\in\A$, 
and if
\begin{equation}\label{eq:identification function}
\E_F \left[ V(Y;a) \right] =0 \quad \Longleftrightarrow  \quad a\in A(F),
\end{equation}
for all $Y \sim F \in {\cal F}$ and for all $a \in \A$.
\end{defi}
In estimation contexts, identification functions can be used for $Z$-estimation or the (generalized) method of moments \cite{Hansen1982, NeweyMcFadden1994} as well as for forecast validation or calibration tests as detailed in \cite{NoldeZiegel2017}.

Let $\F_{\rm cont}^\tau$ be the class of distribution functions on $\R$ which are continuously differentiable and whose $\tau$-quantiles are singletons.
Lemma \ref{lem:ES} yields the following `natural' strict $\F_{\rm cont}^\tau$-identification function for $(\ES_\tau^-, q_\tau, \ES_\tau^+)$
\begin{equation}
\label{eq:V}
V(y;e^-,v,e^+) = 
\begin{pmatrix}
e^- + \tfrac{1}{\tau}S_\tau^-(y;v) \\
\one_{\{y\le v\}} - \tau\\
e^+ - \tfrac{1}{1-\tau}S_\tau^+(y;v)
\end{pmatrix}.
\end{equation}
Clearly, a linear transformation of full rank of a strict identification function is again a strict identification function (and according to \cite[Theorem S.1]{DimitriadisFisslerZiegel2020} there are no further choices, subject to regularity conditions).
Therefore, another natural alternative is 
\begin{equation}
\label{eq:V2}
\widetilde V(y;e^-,v,e^+) = 
\begin{pmatrix}
\tau e^- + (1-\tau)e^+ - y \\
\one_{\{y\le v\}} - \tau\\
e^+ - \tfrac{1}{1-\tau}S_\tau^+(y;v)
\end{pmatrix}
=
\begin{pmatrix}
\tau & 0 & (1-\tau) \\
0 & 1 & 0 \\
0 & 0 & 1
\end{pmatrix}
V(y;e^-,v,e^+).
\end{equation}

\subsection{Assumptions}

The proof of Theorem \ref{thm:necessary} exploits Osband's principle in its form of \cite[Theorem 3.2]{FisslerZiegel2016}. Therefore, it uses a similar set of assumptions, which are also in corresponding results such as \cite[Theorem 3.7]{FisslerZiegel2021}.
More details about their implications and interpretations can be found in \cite{FisslerZiegel2016}.
In the sequel, we use the following shorthand notations
\[
\bar L(F;a) := \E_F[L(Y;a)], \qquad \bar V(F;a) := \E_F[V(Y;a)].
\]

\begin{ass}\label{ass:V1}
$\F$ is convex and for every $a\in\interior(\A)$, the interior of $\A$, there are $F_1, \ldots, F_{4}\in\F$ such that 
\(
0\in \interior\left( \conv\left(\left\{ \bar V(F_1;a), \ldots, \bar V(F_{4};a)\right\}\right)\right)\,.
\)
\end{ass}
Since $V$ is a strict $\F$-identification function for $(\ES_\tau^-, q_\tau, \ES_\tau^+)$, Assumption \ref{ass:V1} implies that $(\ES_\tau^-, q_\tau, \ES_\tau^+)$ maps $\F$ surjectively to $\interior(\A)$.


\begin{ass}\label{ass:V4}
For all $v_0\in  \{v\in\R: \text{there is }(e^-, e^+)\in\R^2 \text{ such that } \ (e^-,v,e^+) \in\A\}$ there are $F_1,F_2 \in \F$ with derivatives $f_1,f_2$ such that $q_\tau(F_1) = q_\tau(F_2) = v_0$ and
$f_1(v)\neq f_2(v)$.
\end{ass}

\begin{ass}\label{ass:F1}
For every $y\in\R$ there exists a sequence $(F_n)_{n \in \mathbb{N}}$ of distributions $F_n \in \F$ that converges weakly to the Dirac-measure $\delta_y$ such that the support of $F_n$ is contained in a compact set $K$ for all $n$. 
\end{ass}

\begin{ass}\label{ass:VS1}
$L$ is locally bounded and 
the complement of the set
\[
C := \{(y;a) \in \R\times \A\;|\; \text{$L(\cdot;a)$ is continuous at the point $y$}\}
\]
has $4$-dimensional Lebesgue measure zero.
\end{ass}

\begin{ass}\label{ass:S2}
For every $F\in \F$, the function $\bar L(F; \cdot)$ is twice continuously differentiable. 
\end{ass}

\subsection{Proof of Theorem \ref{thm:necessary}}
\label{subsec:proof}

The proof exploits Osband's principle \cite[Theorem 3.2]{FisslerZiegel2016}. We use the identification function $V$ given in \eqref{eq:V} rather than the one in \eqref{eq:V2} since this choice is symmetric in the lower and upper ES arguments. 
Let $F\in\F$ with continuous derivative $f$ (which then coincides with one version of the Lebesgue density of $F$). We obtain
\begin{align*}
\bar V_1 (F;e^-,v,e^-) &= e^- + \tfrac{1}{\tau}(F(v) - \tau)v - \tfrac{1}{\tau}\int_{-\infty}^v yf(y) \mathrm{d} y,\\
\bar V_2 (F;e^-,v,e^-) &= F(v) - \tau,\\
\bar V_3 (F;e^-,v,e^-) &= e^+ - \tfrac{1}{1-\tau}(F(v) - \tau)v + \tfrac{1}{1-\tau}\int_{v}^{\infty} yf(y) \mathrm{d} y.
\end{align*}
The non-vanishing partial derivatives of $\bar V(F;\cdot)$ are
\begin{align*}
\partial_{e^-} \bar V_1 (F;e^-,v,e^-) &= 1,\\
\partial_{v} \bar V_1 (F;e^-,v,e^-) &=  \tfrac{1}{\tau}(F(v) - \tau),\\
\partial_v \bar V_2 (F;e^-,v,e^-) &= f(v),\\
\partial_{v} \bar V_3 (F;e^-,v,e^-) &=  -\tfrac{1}{1-\tau}(F(v) - \tau),\\
\partial_{e^+}\bar V_3 (F;e^-,v,e^-) &= 1.
\end{align*}
Osband's principle \cite[Theorem 3.2]{FisslerZiegel2016} yields that, under Assumptions \ref{ass:V1} and \ref{ass:S2}, there are continuously differentiable functions $h_{ij}\colon\interior(\A)\to\R$, $i,j=1, 2, 3$, such that for all $F\in\F$ and all $(e^-,v,e^+)\in \interior(\A)$
\[
\partial_m \bar L(F;e^-,v,e^+) = \sum_{i=1}^3 h_{mi} (e^-,v,e^+) \bar V_i(F;e^-,v,e^+).
\]
Due to Assumption \ref{ass:S2} the Hessian of $\bar L(F;\cdot)$ must be symmetric for any $F\in\F$. This gives us three conditions which provide a lot of information about the matrix $h = (h_{ij})$. For a given distribution $F\in\F$ we shall evaluate these symmetry conditions at the true functional value $(e^-,v,e^+) = (\ES_\tau^-(F), q_\tau(F), \ES_\tau^+(F))$ as well as at an arbitrary value in $\interior(\A)$.

Exploiting that the expected identification function vanishes at the true functional value, we obtain for $\partial_{e^+}\partial_{e^-} \bar L(F,e^-,v,e^+) = \partial_{e^-}\partial_{e^+} \bar L(F,e^-,v,e^+)$ at
$(e^-,v,e^+) = (\ES_\tau^-(F), q_\tau(F), \ES_\tau^+(F))$
\[
h_{13}(e^-,v,e^+) = h_{31}(e^-,v,e^+).
\]
Since we can repeat this argument for any distribution $F\in\mathcal F$ and by exploiting the surjectivity condition, we get 
\begin{equation}
\label{eq:sym0}
h_{13}\equiv h_{31}.
\end{equation}
Considering $\partial_{v}\partial_{e^-} \bar L(F,e^-,v,e^+) = \partial_{e^-}\partial_{v} \bar L(F,e^-,v,e^+)$ at
$(e^-,v,e^+) = (\ES_\tau^-(F), q_\tau(F), \ES_\tau^+(F))$ we get
\[
h_{12}(e^-,v,e^+)f(v) = h_{21}(e^-,v,e^+).
\]
Assumption \ref{ass:V4} together with a surjectivity argument yields
\[
h_{12}\equiv h_{21}\equiv 0.
\] 
With similar arguments, we get
\[
h_{32}\equiv h_{23}\equiv 0.
\] 
For $F_1\in\F$
the condition $\partial_v \partial_{e^-} \bar L(F_1;\cdot) = \partial_{e^-}\partial_v\bar L(F_1;\cdot)$ evaluated for a general point $(e^-,v,e^+)$ implies that 
\begin{align*}
0&=\partial_v h_{11}(e^-,v,e^+)\bar V_1(F_1;e^-,v,e^+) \\
&+\big[\tfrac{1}{\tau} h_{11}(e^-,v,e^+)-\tfrac{1}{1-\tau} h_{13}(e^-,v,e^+) - \partial_{e^-}h_{22}(e^-,v,e^+)\big] \bar V_2(F_1;e^-,v,e^+)\\
&+ \partial_v h_{13}(e^-,v,e^+)V_3(F_1;e^-,v,e^+).
\end{align*}
Assumption \ref{ass:V1} implies that there are $F_2,F_3\in\F$ such that $\bar V(F_i; e^-,v,e^+)$, $i=1,2,3$, are linearly independent. 
Exploiting the surjectivity once again therefore yields
\begin{equation}
\label{eq:sym1}
\partial_v h_{11} \equiv 0, \qquad \tfrac{1}{\tau} h_{11}-\tfrac{1}{1-\tau} h_{13} \equiv \partial_{e^-}h_{22}, \qquad \partial_v h_{13}\equiv0.
\end{equation}
Similarly, the condition $\partial_v \partial_{e^+} \bar L(F;\cdot) = \partial_{e^+}\partial_v\bar L(F;\cdot)$ implies that
\begin{equation}
\label{eq:sym2}
\partial_v h_{31} \equiv 0, \qquad \tfrac{1}{\tau} h_{31}-\tfrac{1}{1-\tau} h_{33} \equiv \partial_{e^+}h_{22}, \qquad \partial_v h_{33}\equiv0.
\end{equation}
The third symmetry condition and a repetition of the same arguments finally yields
\begin{equation}
\label{eq:sym4}
\partial_{e^+} h_{11} \equiv \partial_{e^-} h_{31}, \qquad \partial_{e^-} h_{33} \equiv \partial_{e^+} h_{31}.
\end{equation}
Equations \eqref{eq:sym1} and \eqref{eq:sym2} yield that the functions $h_{11}, h_{33}, \partial_{e^-}h_{22}, \partial_{e^+}h_{22}$ only depend on $(e^-, e^+)$ and are independent of $v$.
Equations \eqref{eq:sym0} and \eqref{eq:sym4} yield that there is a thrice continuously differentiable function $\Phi$ in $(e^-, e^+)$ with Hessian $(h_{ij})_{i,j=1,3}$.
Moreover, equations \eqref{eq:sym1} and \eqref{eq:sym2} imply that there is a function $\eta$ in $v$ such that 
\[
h_{22}(e^-,v,e^+)  = \eta(v) + \tfrac{1}{\tau}\partial_{e^-}\Phi(e^-, e^+)  - \tfrac{1}{1-\tau} \partial_{e^+}\Phi(e^-, e^+).
\]

Due to the strict consistency of $L$, the Hessian of its expectation needs to be positive semi-definite at $(e^-,v,e^+) = (\ES_\tau^-(F), q_\tau(F), \ES_\tau^+(F))$.
This yields that the submatrix $(h_{ij})_{i,j=1,3}$ is positive semi-definite and that $h_{22}\ge0$.

Finally, an application of \cite[Proposition 1]{Erratum} yields the claim, noting that $\interior(\A)$ is simply connected due to the mixture-continuity of $(\ES_\tau^-, q_\tau, \ES_\tau^+)$ and Assumption \ref{ass:V1}.
Note that $g'=\eta$ and that the Hessian of $\Phi$ is $(h_{ij})_{i,j=1,3}$.
\hfill {\scriptsize $\square$}


\newpage
\section{Supplementary material: {\sf R} code}

\begin{lstlisting}[float=h!,frame=tb,caption={Deep multiple quantile regression: additive approach.},label=QuantAdd]
Design  <- layer_input(shape = c(4), dtype = 'float32') 
#
Cat1  <- layer_input(shape = c(1), dtype = 'int32')
Cat1Emb = Cat1 %>% 
      layer_embedding(input_dim=24, output_dim=2, input_length=1) %>% layer_flatten()
#
Cat2  <- layer_input(shape = c(1), dtype = 'int32')
Cat2Emb = Cat2 %>% 
      layer_embedding(input_dim=13, output_dim=2, input_length=1) %>% layer_flatten()
#
Cat3  <- layer_input(shape = c(1), dtype = 'int32')
Cat3Emb = Cat3 %>% 
      layer_embedding(input_dim=16, output_dim=2, input_length=1) %>% layer_flatten()
#
Network = list(Design, Cat1Emb, Cat2Emb, Cat3Emb) %>% layer_concatenate() %>%
            layer_dense(units=20, activation='tanh', name='FNLayer1') %>%
            layer_dense(units=15, activation='tanh', name='FNLayer2') %>%
            layer_dense(units=10, activation='tanh', name='FNLayer3') 
#
Q1  <- Network %>% layer_dense(units=1, activation='exponential')
#
Q20 <- Network %>% layer_dense(units=1, activation='exponential')
Q2  <- list(Q1,Q20) %>% layer_add()
#
Q30 <- Network %>% layer_dense(units=1, activation='exponential')
Q3  <- list(Q2,Q30) %>% layer_add()
#
model <- keras_model(inputs = list(Design), outputs = c(Q1,Q2,Q3))


\end{lstlisting}

\begin{lstlisting}[float=h!,frame=tb,caption={Deep multiple quantile regression: multiplicative approach.},label=QuantMult]
Q3  <- Network %>% layer_dense(units=1, activation='exponential')
#
Q20 <- Network %>% layer_dense(units=1, activation='sigmoid')
Q2  <- list(Q3,Q20) %>% layer_multiply()
#
Q10 <- Network %>% layer_dense(units=1, activation='sigmoid')
Q1  <- list(Q2,Q10) %>% layer_multiply()


\end{lstlisting}

\begin{lstlisting}[float=h!,frame=tb,caption={Fitting a deep multiple quantile regression
with pinball losses.},label=QuantFitting]
PinB1 <- function(y_true, y_pred){k_mean(k_maximum(y_true - y_pred, 0) * 0.1 
                                    + k_maximum(y_pred - y_true, 0) * (1 - 0.1))}
PinB2 <- function(y_true, y_pred){k_mean(k_maximum(y_true - y_pred, 0) * 0.5 
                                    + k_maximum(y_pred - y_true, 0) * (1 - 0.5))}
PinB3 <- function(y_true, y_pred){k_mean(k_maximum(y_true - y_pred, 0) * 0.9 
                                    + k_maximum(y_pred - y_true, 0) * (1 - 0.9))}
# 
model %>% compile(loss = list(PinB1, PinB2, PinB3), optimizer = 'nadam')
\end{lstlisting}

\begin{lstlisting}[float=h!,frame=tb,caption={Fitting a deep composite model regression
with scoring function \eqref{eq:score revelation}.},label=CompositeFitting]
Loss <- function(y_true, y_pred){
       mu <- tau0 * y_pred[,1] + (1-tau0) * y_pred[,3]
       k_mean( 
         (k_maximum(y_true[,1] - y_pred[,2], 0) * 0.9 + 
                        k_maximum(y_pred[,2]- y_true[,1], 0) * 0.1)*
           (1 - c2 * y_pred[,3]^(b2-1) / ((b2-1) * 0.1)) +
         c2/(b2-1) * (y_pred[,3]^(b2-1)*(y_pred[,3]-y_true[,1])
                         -y_pred[,3]^b2/b2+y_true[,1]^b2/b2)+
         c1/(b1-1) * (mu^(b1-1)*(mu_true[,1])-mu^b1/b1+y_true[,1]^b1/b1)) }
#
model %>% compile(loss = Loss, optimizer = 'nadam')
\end{lstlisting}


\begin{thebibliography}{999}


\bibitem{Barendse2020}
Barendse, S. (2020). 
Efficiently weighted estimation of tail and interquartile expectations. 
{\it SSRN Manuscript} ID 2937665.



\bibitem{CoorayAnanda}
Cooray, K., Ananda, M.M.A. (2005).
Modeling actuarial data with composite lognormal-Pareto model. 
{\it Scandinavian Actuarial Journal} {\bf 2005/5}, 321--334.

\bibitem{DimitriadisBayer}
Dimitriadis, T., Bayer, S. (2019).
A joint quantile and expected shortfall regression framework.
{\it Electronic Journal of Statistics} {\bf 13/1}, 1823--1871.


\bibitem{DimitriadisFisslerZiegel2020}
Dimitriadis, T., Fissler, T., Ziegel, J.F. (2020).
The efficiency gap.
{\it arXiv}, 2010.14146.




\bibitem{EmbrechtsWang2015}
Embrechts, P., Wang, R. (2015).
Seven proofs for the subadditivity of expected shortfall.
{\it Dependence Modeling} {\bf 3}, 126--140.


\bibitem{FisslerZiegel2016}
Fissler, T., Ziegel, J.F. (2016).
Higher order elicitability and Osband's principle.
{\it The Annals of Statistics} {\bf 44/4}, 1680--1707.

\bibitem{Erratum}
Fissler, T., Ziegel, J.F. (2021).
Correction note: Higher order elicitability and Osband's principle.
{\it The Annals of Statistics} {\bf 49/1}, 614.


\bibitem{FisslerZiegel2021}
Fissler, T., Ziegel, J.F. (2021).
On the elicitability of range value at risk.
{\it Statistics \& Risk Modeling}  {\bf 38/1--2}, 25--46.


\bibitem{FrongilloKash2020}
Frongillo, R., Kash, I. (2021).
\newblock Elicitation complexity of statistical properties.
\newblock \emph{Biometrika} {\bf 108/4}, 857--879.


%
\bibitem{Fung2021}
  Fung, T.C., Badescu, A.L., Lin, X.S. (2021).
  A new class of severity regression models with an application to IBNR prediction.
  {\it North American Actuarial Journal} {\bf 25/2}, 206--231.


\bibitem{GanValdez}
Gan, G., Valdez, E.A. (2018).
Fat-tailed regression modeling with spliced distributions.
{\it North American Actuarial Journal} {\bf 22/4}, 554--573.

\bibitem{Gneiting}
Gneiting, T. (2011).
Making and evaluating point forecasts.
{\it Journal of the American Statistical Association} {\bf 106/494}, 746--762.

\bibitem{Gneiting2011b}
Gneiting, T. (2011).
\newblock Quantiles as optimal point forecasts.
\newblock {\it International Journal of Forecasting}, {\bf 27/2}, 197--207.

\bibitem{GneitingRaftery}
Gneiting, T., Raftery, A.E. (2007).
Strictly proper scoring rules, prediction, and estimation.
{\it Journal of the American Statistical Association} {\bf 102/477}, 359--378.



\bibitem{Grun}
Gr\"un, B., Miljkovic, T. (2019). 
Extending composite loss models using a general framework of advanced
computational tools. 
{\it Scandinavian Actuarial Journal} {\bf 2019/8}, 642--660.

\bibitem{Guillen2021}
Guillen, M., Berm\'udez, L., Pitarque, A., (2021).
Joint generalized quantile and conditional tail expectation for insurance risk analysis.
{\it Insurance: Mathematics and Economics} {\bf 99}, 1--8.

\bibitem{Hansen1982}
Hansen, L.P. (1982).
Large sample properties of generalized method of moments estimators.
{\it Econometrica} {\bf 50/4}, 1029--54.

\bibitem{HTF}
Hastie, T., Tibshirani, R., Friedman, J. (2009).
{\it The Elements of Statistical Learning.
Data Mining, Inference, and Prediction.}
2nd edition. Springer Series in Statistics.


  
  
\bibitem{Jorgensen2}
  J{\o}rgensen, B. (1987).
  Exponential dispersion models.
  {\it Journal of the Royal Statistical Society, Series B} {\bf 49/2},
  127--145.


\bibitem{KoenkerBassett}
Koenker, R., Bassett, G., Jr. (1978).
Regression quantiles.
{\it Econometrica} {\bf 46/1}, 33--50.



\bibitem{Laudage}
Laudag\'e, C., Desmettre, S., Wenzel, J. (2019). 
Severity modeling of extreme insurance claims for tariffication.
{\it Insurance: Mathematics and Economics} {\bf 88}, 77--92.


\bibitem{McNeil}
McNeil, A.J., Frey, R., Embrechts, P. (2015).
{\it Quantitative Risk Management: Concepts, Techniques and Tools.}
Revised edition. Princeton University Press.

\bibitem{Meinshausen}
Meinshausen, N. (2006). 
Quantile regression forests. 
{\it Journal of Machine Learning Research} {\bf 7}, 983--999.



\bibitem{NeweyMcFadden1994}
Newey, W.K., McFadden, D. (1994).
\newblock {L}arge sample estimation and hypothesis testing.
\newblock In R.F. Engle and D. McFadden (Eds.), {\em {Handbook of
  Econometrics}},  Volume~4, Chapter~36, Elsevier, 2111--2245. 
  
 \bibitem{NoldeZiegel2017}
Nolde, N., Ziegel, J.F. (2017).
\newblock {Elicitability and backtesting: Perspectives for banking regulation}.
\newblock {\em Annals of Applied Statistics} {\bf 11/4}, 1833--1874.


\bibitem{Osband1985}
Osband, K.H. (1985).
\newblock {\em {Providing Incentives for Better Cost Forecasting}}.
\newblock PhD thesis, University of California, Berkeley.


\bibitem{Parodi}
Parodi, P. (2020).
A generalised property exposure rating framework that incorporates scale-independent losses
and maximum possible loss uncertainty. 
{\it ASTIN Bulletin} {\bf 50/2}, 513--553.



\bibitem{PigeonDenuit}
Pigeon, M., Denuit, M.. 
Composite lognormal-Pareto model with random threshold. 
{\it Scandinavian Actuarial Journal} {\bf 2011/3}, 177--192.


\bibitem{Richman3}
Richman, R. (2021).
Mind the gap -- safely incorporating deep learning models into the actuarial toolkit.
{\it SSRN Manuscript} ID 3857693.



\bibitem{Saerens}
Saerens, M. (2000).
Building cost functions minimizing to some summary statistics.
{\it IEEE Transactions on Neural Networks}  {\bf 11}, 1263--1271.

\bibitem{Savage}
Savage, L.J. (1971).
Elicitable of personal probabilities and expectations.
{\it Journal of the American Statistical Association} {\bf 66/336}, 783--810.

\bibitem{Scollnik}
Scollnik, D.P.M. (2007).
On composite lognormal-Pareto models. 
{\it Scandinavian Actuarial Journal} {\bf 2007/1}, 20--33.


\bibitem{Takeuchi}
Takeuchi, I., Le, Q.V., Sears, T.D., Smola, A.J. (2006). 
Nonparametric quantile estimation.
{\it Journal of Machine Learning Research} {\bf 7}, 1231--1264.


\bibitem{Thomson}
Thomson, W. (1979). 
Eliciting production possibilities from a well-informed manager
{\it Journal of Economic Theory} {\bf 20}, 360--380.

\bibitem{Tweedie}
Tweedie, M.C.K. (1984).
An index which distinguishes between some important exponential families. 
In: {\it Statistics: Applications and New Directions}. 
Ghosh, J.K., Roy, J. (Eds.).
Proceeding of the Indian Statistical Golden Jubilee International Conference, Indian Statistical Institute, Calcutta, 579--604.


\bibitem{UribeGuillen}
Uribe, J.M., Guillen, M. (2019).
{\it Quantile Regression for Cross-Sectional and Time Series Data Applications in Energy
Markets using {\sf R}}. Springer.

\bibitem{VanderVaart2}
Van der Vaart, A.W. (1998). 
{\it Asymptotic Statistics.} Cambridge University Press.


\bibitem{Weber2006}
Weber, S. (2006). 
Distribution-invariant risk measures, information, and dynamic consistency.
{\it Mathematical Finance} {\bf 16}, 419--441.


\bibitem{WM2021}
W\"uthrich, M.V., Merz, M.  (2021).
Statistical foundations of actuarial learning and its applications.
{\it SSRN Manuscript} ID 3822407.


\end{thebibliography}
\end{document}